\documentclass[11pt]{article}
\usepackage{epstopdf}                                                          
\usepackage[a4paper,hmarginratio=1:1,vmarginratio=2:3,totalwidth=15.2cm,totalheight=22.77cm]{geometry}
\usepackage{bm,epstopdf,epsfig,amsmath,amssymb,
amsfonts,colordvi,wrapfig,comment,cancel,verbatim,slashed}
\usepackage{ulem}
\usepackage{graphicx,graphics}
\usepackage[font=md,captionskip=8pt]{subfig}
\usepackage[usenames,dvipsnames]{color}
\usepackage[noadjust]{cite}
\usepackage{xcolor} 
\usepackage[utf8]{inputenc}
\usepackage{setspace}

\usepackage[utf8]{inputenc}

\newcommand{\w}{\omega}  
\newcommand{\Om}{\omega}

\allowdisplaybreaks

\newcommand{\tr}{\text{tr}}

\newcommand{\be}{\begin{equation}}
\newcommand{\ee}{\end{equation}}
\newcommand{\bea}{\begin{eqnarray}}
\newcommand{\eea}{\end{eqnarray}}
                
\newcommand{\ra}{\rightarrow}  
\newcommand{\Ra}{\Rightarrow}

\newcommand{\baa}{\begin{array}}
\newcommand{\eaa}{\end{array}}

\long\def\symbolfootnote[#1]#2{\begingroup
\def\thefootnote{\fnsymbol{footnote}}\footnote[#1]{#2}\endgroup}

\begin{document} 
\begin{flushright}
\end{flushright}

\thispagestyle{empty}
\vspace{3.cm}
\begin{center}

   {\Large \bf Conformal geometry as a gauge theory of gravity:}
\bigskip

  {\Large\bf covariant  equations of motion \& conservation laws}

 \vspace{1.5cm}
 
 {\bf  C. Condeescu},\,\,  {\bf  D. M. Ghilencea}\,\, and  {\bf  A. Micu}\,\,
 \symbolfootnote[1]{E-mail: ccezar@theory.nipne.ro, dumitru.ghilencea@cern.ch, amicu@theory.nipne.ro}
 
\bigskip\bigskip

{\small Department of Theoretical Physics, National Institute of Physics
\smallskip 

 and  Nuclear Engineering (IFIN), Bucharest, 077125 Romania}
\end{center}

\medskip

\begin{abstract}
   \noindent 
  We study Weyl conformal geometry as a general gauge theory of the Weyl group
  (of Poincar\'e and dilatations symmetries)  in a manifestly Weyl gauge
  covariant formalism in which this geometry   is automatically metric and
  physically relevant. This gives a realistic (quadratic) gauge theory of
  gravity, with  Einstein-Hilbert gravity  recovered in its spontaneously
  broken phase, motivating  our interest in this geometry.
 For the most general action we compute the manifestly Weyl gauge covariant
  equations of motion  and present the conservation laws for the
energy-momentum tensor and Weyl gauge current. These laws are  valid
both in Weyl conformal geometry (with respect to the Weyl gauge covariant derivative)
but also in the Riemannian geometry equivalent picture (with respect to its
associated covariant derivative). This interesting result is a
consequence of gauged diffeomorphism invariance of the former versus
usual diffeomorphism invariance of the latter. These results are first derived
in $d=4$ dimensions. We then successfully derive the conservation laws and
equations of motion in Weyl conformal geometry in arbitrary $d$ dimensions, while
maintaining manifest Weyl gauge invariance/covariance. The results are useful in
physical applications with this symmetry.
\end{abstract}

\newpage
$ $
\vspace{4cm}
\tableofcontents{}

\newpage

\section{Motivation}\label{1}

Much effort has been devoted to find a unified ultraviolet complete theory of
gravity and Standard Model of particle theory. String theory is one
possible solution to this problem. Here we consider  an alternative
approach, represented by the (quantum)  gauge theory of gravity {\it defined}
by Weyl conformal geometry  \cite{Weyl1,Weyl2,Weyl3},  see \cite{review} for a review.

Weyl conformal geometry, introduced in 1918, long before modern gauge theories -
was in essence  the first ever  gauge theory (of a space-time symmetry).
These days it can be introduced and motivated
by the successful principle of gauge symmetries published  by
 E. Noether \cite{Noether} in the same year; much like searches
for new physics based on new internal gauge symmetries, one can apply this
principle to  a new 4D space-time symmetry to identify the 4D geometry
that may underlie both gravity and SM in a (unified) gauge theory.
The gauged space-time symmetry considered  dictates the actual 4D geometry
and the gravity action {\it as a gauge theory action}. To this purpose, one can
gauge: a) Poincar\'e group \cite{Utiyama,Kibble},
b) Weyl  group (Poincar\'e and dilatations symmetry)
\cite{Tait,Bregman,CDA2}, c)  full conformal group \cite{Freedman}.

The action in case a) has a largely  unknown UV completion, being affected by
an infinite series of higher derivative operators, the presence of which
cannot be forbidden on symmetry grounds. We do not consider this case.
Case c) of gauging the full conformal group leads to conformal gravity \cite{mannheim}.
This case is {\it not}  a true gauge theory of a space-time symmetry
since {\it no dynamical/physical} gauge bosons of dilatations and special conformal
symmetries can be present in the  action \cite{Kaku}. This leads us to explore
case b) of the gauge theory of the (smaller) Weyl group. This symmetry is realised by
Weyl conformal geometry \cite{Weyl1,Weyl2,Weyl3} that was  less studied as a gauge theory; for a  modern gauge theory view  see \cite{CDA2}.

The reason why conformal geometry as a gauge theory of Weyl group has not been well
studied  is partly due to a century-old criticism of Einstein
\cite{Weyl1}\footnote{This
was in response to the (wrong) initial idea of Weyl that
the Weyl gauge boson of dilatations ($\w_\mu$)
in conformal geometry would describe the real photon.
But Einstein's critique is more general and applies independent of this idea.
It was much later understood that electromagnetism corresponds  to an internal,
(not a space-time) symmetry. The vector field $\omega_\mu$ is thus just a field that
together with $g_{\mu\nu}$  mediates gravity.}.
According to this, a theory based on conformal geometry cannot be physical since this
geometry is non-metric, leading to a 
``second clock effect'' that contradicts experiment \cite{Weyl1}.
Therefore,  research was focused on theories based on
so-called ``integrable''  conformal geometry, in which the Weyl gauge
boson of dilatations is ``pure gauge'' (non-dynamical); in this case
conformal geometry becomes ``metric'', avoids the above criticism, and one
obtains the conformal gravity action \cite{Kaku} (which we just
argued that it is {\it not} a true gauge theory of a space-time symmetry). 
Conformal geometry remained however of interest to physics, 
and  set the foundations for modern gauge theories, see  \cite{Scholz}
and references therein. With important
exceptions  \cite{Dirac,Smolin},  with the advent of modern gauge theories
(of internal symmetries)  attention was diverted to other emerging theories
at the time (SM as a gauge theory, supersymmetry, string theory).

However, the criticism of Einstein does not actually apply to the gravity theory
of conformal geometry. Recently, it was shown \cite{Ghilen0}  that the general {\it quadratic}
gravity action associated to conformal geometry undergoes a Stueckelberg
mechanism \cite{Stueckelberg}. In this, the Weyl gauge boson ($\omega_\mu$)
of dilatations becomes
massive,  after eating the would-be-Goldstone of dilatations $\phi$ (a spin-zero state
propagated by the $R^2$ term in the action) and then  Weyl gauge symmetry is
broken. After $\omega_\mu$ decouples,  Weyl geometry (i.e. connection)
becomes Riemannian (Levi-Civita connection) and Einstein-Hilbert gravity is recovered, with
a  cosmological constant ($\Lambda>0$). All mass scales of
this  theory i.e. Planck, $m_\omega$ and $\Lambda$ have a geometric origin, generated
by the would-be-Goldstone vev (``dilaton''). Since this field is
 eaten by $\omega_\mu$, there is no need to ``stabilize'' this value.
No scalar fields were added to the theory to generate these scales
and the breaking has a pure geometric origin  \cite{Ghilen0,non-metricity}.

After $\omega_\mu$ decouples, any unwanted non-metricity effects that one may claim
to exist can be suppressed by a sufficiently large  mass of $\omega_\mu$
(that could be e.g. near Planck scale) \cite{Ghilen0}.  It was also
subsequently realised that if Weyl gauge covariance is indeed respected by the
parallel transport of the vector, its length (or clock rates) is invariant and therefore the
geometry is actually {\it metric} \cite{Lasenby,non-metricity,CDA}, with
no unwanted second-clock effect present (in the symmetric phase).
(Another argument is that in the symmetric phase there is no mass scale
and without a mass scale  a clock rate cannot be defined and thus no second
clock effect exists).
These two developments
built on previous works of Dirac \cite{Dirac} and Smolin \cite{Smolin} and 
showed that conformal geometry, regarded as a gauge theory of Weyl group,
can be a realistic vector-tensor gauge theory of gravity, called here
Weyl quadratic gravity.

Another motivation for an interest in conformal geometry as a gauge theory
is that this theory is Weyl (gauge) anomaly-free. This was shown in \cite{DG1}
by introducing a Weyl gauge covariant formulation in which this geometry is
automatically {\it metric.}
While in Riemannian geometry a  Weyl anomaly is present in  an action
with {\it local} Weyl symmetry   \cite{Duff,Duff2,Duff3,Deser1976,Deser},
in conformal geometry the Weyl gauge symmetry of dilatations is maintained
and manifest in $d$ dimensions and thus at quantum level \cite{DG1,review}.
Weyl geometry is special in that  both the Weyl {\it and}
Chern-Euler-Gauss-Bonnet terms in the action transform Weyl gauge covariantly in $d$ dimensions
(unlike the Chern-Euler-Gauss-Bonnet term in Riemannian case).
As a result we have a (geometric) regularisation and action in
$d$ dimensions that are Weyl gauge invariant and thus Weyl anomaly is absent.
The absence of anomaly is also  due to the presence of an extra degree of freedom
(``dilaton'' or would-be-Goldstone) \cite{Englert}. Weyl anomaly is recovered
in the  broken phase when $\omega_\mu$ decouples
and Weyl geometry (connection) becomes Riemannian (Levi-Civita).
Briefly, conformal geometry is a consistent (quantum) gauge theory of the Weyl group
and {\it the only true gauge theory} of a space-time symmetry i.e. with physical (dynamical)
$\w_\mu$ and  full geometric interpretation \cite{CDA}.

There are additional motivations to study conformal geometry as the
gauge theory beyond Einstein-Hilbert gravity.
The Standard Model (SM) with vanishing Higgs mass parameter
can naturally  be embedded in conformal geometry \cite{SMW} with no additional
degrees of freedom (dof) beyond those of SM and Weyl geometry ($\w_\mu$, $g_{\mu\nu}$).
Successful Starobinsky-Higgs inflation is
possible \cite{WI3,WI1,WI2} being a gauged version of
Starobinsky inflation \cite{Starobinsky} - this is because
the action is quadratic in curvature (as in any gauge theory) which
 explains the  Starobinsky-like inflation.
Good fits for  galaxies rotation curves seem possible \cite{Harko}
with $\w_\mu$ as a candidate for dark matter; with $\omega_\mu$ of geometric origin,
this gives a geometric solution to this problem from the Weyl geometry
perspective, while in the Riemannian picture $\omega_\mu$ appears
as an ad-hoc  ``matter'' dof; black hole solutions were studied in \cite{Harko2}.

Interestingly, there exists
a non-perturbative  completion of Weyl quadratic gravity action in a more general
gauge theory of the Weyl group, having a non-linear  action.
This is  an analogue of the  Dirac-Born-Infeld  action
\cite{D,BI,Sorokin,Gibbons} associated to the gauged dilatation of  conformal geometry
in $d$ dimensions (hereafter Weyl-DBI) \cite{WDBI}. This action is itself a 
theory that  is Weyl  gauge invariant in arbitrary  $d$ dimensions  with
{\it dimensionless} couplings and no UV regulator needed.
This is also true for Weyl quadratic gravity after  a
Weyl gauge invariant regularisation
with (covariant) scalar curvature $R$ acting as UV regulator.
The Weyl-DBI action is a sum of two terms:  a  perturbative gauge structure which is 
 Weyl quadratic gravity and a non-perturbative part of
(gauge invariant) non-polynomial operators. Explicitly,
a series expansion of this WDBI action
recovers in the leading order exactly the (gauge theory of) Weyl quadratic
gravity in $d$ dimensions of conformal geometry, while all sub-leading
terms in such expansion are    non-perturbative (quantum) corrections.

Recently, it was  found that there are three mathematically equivalent
formulations  \cite{CDA,CDA2} of conformal geometry: 1) a ``traditional''
historical formulation that has vectorial non-metricity but is torsion-free;
2) a second, equivalent  formulation that is metric but has vectorial
torsion, derived  from the tangent space formulation
uplifted to space-time, and finally, 3) a third
formulation that is Weyl gauge covariant and automatically metric
\cite{DG1,CDA,CDA2}.  These formulations
correspond to different linear combinations of the Weyl group generators \cite{CDA2}.
In general, one can use any  formulation, but only
the third formulation is guaranteed to be physical,
because conformal geometry is a gauge theory of the Weyl group, hence
Weyl gauge covariance must be respected.
Given the underlying symmetry, we can  say that the third
formulation of  conformal geometry is simply a {\it  ``covariantised''
version of Riemannian geometry} with respect to the gauged dilatation
symmetry.

The goal of this paper is three-fold: first, we review and complete
the  Weyl gauge covariant mathematical formalism
\cite{Dirac,DG1,CDA,CDA2}. Second,
for the most general  action of this gauge theory (i.e. Weyl quadratic gravity)
in $d\!=\!4$ dimensions
we find the equations of motion, using the Weyl gauge covariant {\it metric}
(non-affine!) formulation of Weyl
conformal geometry. These equations are manifestly Weyl gauge covariant and can be used in
applications.
Previous work  considered only a non-covariant, non-metric formulation, for simpler actions;
such equations were  obtained by going to the (metric) Riemannian picture
 \cite{SMW,SMW2}.
We also find the conservation laws for the general quadratic gravity action.
We then  extend  these results to  arbitrary $d$ dimensions, respecting
Weyl gauge invariance of\,the\,action.

The conservation laws that we find
for the Weyl gauge current and the stress energy tensor
are formulated in a Weyl gauge covariant form of  conformal geometry.
Then they are shown to also be  valid in the Riemannian geometry in their corresponding
covariant formulation.
It is very interesting to find that these  laws are valid
both in Weyl and Riemannian geometry pictures  of the same action,
with respect to different covariances (Weyl versus Riemannian)!
This interesting result is a consequence of having, separately, standard diffeomorphism
invariance (and dilatation invariance) in the  Riemannian picture
versus gauge covariant diffeomorphisms in conformal geometry formulation.
These conservation laws show the physical relevance of
conformal  geometry as a realistic gauge theory of gravity. These symmetry arguments
are independent of the number of dimensions, which explains why these
conservation laws are also valid in  arbitrary $d$  dimensions.

The above results on  Weyl quadratic gravity with an underlying conformal geometry
endorse  the view that this is a realistic
gauge theory of the Weyl group. Both the geometry
(connection) and the action are Weyl gauge invariant, something
not realised in any other gauge theory of a space-time symmetry.
In the absence of matter, the theory has no parameters other
than the dimensionless couplings of  the action (dictated
by Weyl gauge symmetry alone). The theory is  predictive: it has no
scalar fields compensators added ad-hoc; all dof's are exactly those of conformal
geometry. Thus, we have a geometry constructed by the genius of Weyl that has, at the
same time, all the elegant attributes (general gauge covariance, conservation laws)
of a  realistic gauge theory beyond Einstein-Hilbert gravity, in which an
embedding of the SM (with no new dof's) is naturally realised.

\section{Conformal geometry in the gauge covariant formulation} 
\label{2}

In this section we review the manifestly Weyl gauge covariant formulation of
conformal geometry and of its associated  action,  Weyl quadratic
gravity, regarded  {\it as a gauge theory} of the Weyl group (of Poincar\'e and dilatation
space-time symmetries). This is similar to building an action with an  internal
gauge symmetry,  based on the Noether principle.

Weyl gauge covariance is important firstly because it ensures that the approach
has physical relevance and secondly, the geometry also becomes metric \cite{DG1}.
For later use, we review and complete this formulation of conformal geometry, following
\cite{DG1,CDA,CDA2}. The standard,  modern gauge theories
approach  is to make use of the Lie algebra of the Weyl group and  construct
the theory  on the tangent space, then ``uplift'' the results to curved
space-time, with the vielbein. We briefly present this approach in the Appendix,
for details see \cite{CDA,CDA2}. 
Here we shall  present the gauge covariant formulation of
Weyl conformal geometry directly in space-time, using the more familiar, historic
perspective. The results of this section are valid in arbitrary  $d$ dimensions\footnote{
For Riemannian geometry our conventions in this paper  are as in \cite{Buchbinder}.}.

\bigskip\noindent
{\bf $\bullet$ Non-metric, torsion-free (affine) formulation:}

\bigskip\noindent
Conformal geometry is usually defined by equivalence classes of the metric $g_{\mu\nu}$ and
Weyl gauge field  $\omega_\mu$ related by  Weyl gauge symmetry transformations
defined by $\Sigma(x)\! \equiv\! e^{\lambda_D(x)}$:
\begin{align}
g_{\mu \nu} & \mapsto e^{2 \lambda_D }g_{\mu \nu}\, , & \w_\mu \mapsto \w_\mu - \partial_\mu \lambda_D \, ,
\label{gauge-transformation}
\end{align}

\medskip\noindent
together with the so-called ``non-metricity'' condition of conformal geometry:
\medskip
\bea\label{def}
\tilde \nabla_\mu g_{\nu\rho}=-2 \omega_\mu \,g_{\nu\rho}\, , \qquad \textrm{where}\qquad
\tilde\nabla_\mu g_{\nu\rho}\equiv \partial_\mu g_{\nu\rho}-\tilde\Gamma_{\mu\nu}^\alpha g_{\alpha\rho}
-\tilde\Gamma_{\mu\rho}^\alpha g_{\nu\alpha} \, ,
\eea

\medskip\noindent
with {Weyl connection} $\tilde\Gamma$ assumed symmetric in Weyl conformal geometry:
$\tilde\Gamma_{\mu\nu}^\rho=\tilde\Gamma_{\nu\mu}^\rho$ i.e. there is no torsion.
As shown in (\ref{gauge-transformation}) the metric has Weyl charge $q=2$.
If one would like to  work with an arbitrary charge $q$ of the metric and
also restore the presence of the Weyl gauge coupling $\alpha$,
one replaces  everywhere in the results below $\omega_\mu\ra(\alpha q/2)\,\omega_\mu$.

In this formulation,
one solves (\ref{def}) for $\tilde\Gamma$ by usual techniques, or
by a Weyl gauge covariant derivative replacement:
$\partial_\mu\ra \partial_\mu + 2 \alpha \omega_\mu$ in the Levi-Civita connection, to find
\begin{equation}\label{tildeG}
  \tilde \Gamma_{\mu \nu}^\rho  =\mathring
  \Gamma_{\mu\nu}^\rho\Big\vert_{\partial_\mu\ra \partial_\mu + 2 \alpha \omega_\mu}=
  \mathring \Gamma_{\mu \nu}^\rho + \delta_\mu^\rho \w_\nu  + \delta_\nu^\rho \w_\mu - g_{\mu \nu}\w^\rho\, ,
\end{equation}

\medskip\noindent
with the Levi-Civita connection 
\bea
\mathring \Gamma_{\mu \nu}^\rho = \frac12 g^{\rho \lambda}(\partial_\mu g_{\nu \lambda}
+ \partial_\nu g_{\mu \lambda} - \partial_\lambda g_{\mu \nu}) \, .
\eea

\medskip\noindent
It is easy to see that $\tilde \Gamma$ is invariant under gauged dilatations
(\ref{gauge-transformation}),
  $\tilde \Gamma_{\mu \nu}^{\rho} \mapsto \tilde \Gamma_{\mu \nu}^\rho\,$.

Further, {in this formulation of Weyl geometry}
one  associates  a {Riemann curvature tensor $\tilde R^\rho{}_{\sigma \mu \nu}$
  (hereafter Weyl-Riemann)}  to the Weyl connection $\tilde \Gamma$,
via the commutator of covariant derivatives {($\tilde\nabla$)} acting
on a vector $V^\rho$
\medskip
\begin{equation}\label{t1}
[\tilde \nabla_\mu, \tilde \nabla_\nu] V^\rho = \tilde R^\rho{}_{\sigma \mu \nu} V^\sigma \, .
\end{equation}

\medskip\noindent
From this, one finds for the curvature tensor
an expression similar to that in a Riemannian case but with
the replacement: $\mathring\Gamma_{\mu\nu}^\rho\ra\tilde\Gamma_{\mu\nu}^\rho$:
\medskip
\begin{equation}
\tilde R^\rho{}_\sigma{}_{\mu \nu} = \partial_\mu \tilde \Gamma^\rho_{\nu \sigma}
  - \partial_\nu \tilde \Gamma^\rho_{\mu \sigma} + \tilde \Gamma^\rho_{\mu \tau}
  \tilde \Gamma^\tau_{\nu \sigma} - \tilde \Gamma^\rho_{\nu \tau} \tilde \Gamma^\tau_{\mu \sigma}\, .
\label{curvature-1}
\end{equation} 

\medskip\noindent
Using (\ref{tildeG}), one can further express $\tilde R^\rho{}_{\sigma\mu\nu}$ in terms
of its Riemannian  geometry counterpart\footnote{See later, eqs.(\ref{relation}) to (\ref{riem-3}).} {$(\mathring R^\rho_{\,\,\,\sigma\mu\nu}$)}.
The tensor $\tilde R^\rho{}_{\sigma\mu \nu}$
is invariant under (\ref{gauge-transformation})
since $\tilde\Gamma_{\mu\nu}^\rho$ is so, and the same is true for the Ricci tensor
of Weyl geometry (Weyl-Ricci), $\tilde R_{\sigma\nu}=\tilde R^\rho{}_{\sigma\rho\nu}$,
while the Weyl scalar curvature $\tilde R=g^{\mu\nu} \tilde R_{\mu\nu}$ transforms
Weyl gauge covariantly under (\ref{gauge-transformation}), just like the inverse
metric $\tilde R\ra e^{-2\lambda_D} \tilde R$.  One can also define the field
strength of gauged dilatations 
\medskip
\begin{equation}
  F_{\mu \nu} = \partial_\mu \w_\nu - \partial_\nu \w_\mu  =
  \tilde\nabla_\mu \w_\nu-\tilde\nabla_\nu\w_\mu  \, ,
\label{curvature-3}
\end{equation}

\medskip\noindent
where we used that the connection $\tilde \Gamma_{\mu \nu}^\rho$ is symmetric. $F_{\mu\nu}$
is also invariant under (\ref{gauge-transformation}).
These curvature fields  are then used to construct a Weyl gauge invariant action.
This is done by constructing all quadratic terms invariant under (\ref{gauge-transformation}).
The corresponding action was written by Weyl more than
a century ago \cite{Weyl1,Weyl2,Weyl3} and we shall discuss it shortly.

This picture corresponds to the so-called non-metric {(affine)} formulation. Since the
connection is assumed symmetric, it is also torsion free. This formulation can be
used in applications but since it is non-metric, for calculations one must go
to the (metric) Riemannian picture, see e.g. \cite{SMW}.
Moreover, while the curvature fields
(Weyl-Riemann, Weyl-Ricci, Weyl scalar curvature) are Weyl gauge invariant or transform
covariantly, the  derivatives $\tilde\nabla_\mu$ applied to these curvatures
do {\it not} transform covariantly $\tilde \nabla_\mu \tilde R\not\mapsto
e^{-2\lambda_D}\tilde\nabla_\mu  \tilde R$, etc.  Hence, this is not a suitable
formulation of conformal geometry as a gauge theory of the Weyl group.

\bigskip\noindent
{\bf $\bullet$ Metric {(affine)} formulation with torsion}

\bigskip\noindent
Recently, a different, but equivalent  {(affine)} formulation of Weyl
conformal geometry has been found \cite{CDA} in terms of a new {Weyl} connection $\Gamma$ that
is metric but has vectorial torsion.
Briefly, the idea is that in the non-metricity condition (\ref{def})
$\tilde\nabla_\mu g_{\nu\rho}=-2 \w_\mu g_{\nu\rho}$ one ``moves'' the rhs
into a new connection $\Gamma$,   so that
the new formulation becomes  {\it metric} with respect to $\nabla(\Gamma)$
\medskip
\bea
\nabla_\mu g_{\nu\rho}=0\, ,\qquad \textrm{where}\qquad
  \nabla_\mu g_{\nu\rho}\equiv \partial_\mu g_{\nu\rho} -
\Gamma_{\mu\nu}^\alpha g_{\alpha\rho}-\Gamma_{\mu\rho}^\alpha g_{\alpha\nu} \, .
\eea
This is possible for
 \bea\label{projective}
   \Gamma_{\mu \nu}^\rho =\tilde\Gamma_{\mu\nu}^\rho-\w_\mu \,\delta_\nu^\rho \, ,
   \eea
   or, using (\ref{tildeG})
   \bea\label{rel}
   \Gamma_{\mu\nu}^\rho
=   \mathring \Gamma_{\mu \nu}^\rho + \delta_\mu^\rho \w_\nu - g_{\mu \nu}\w^\rho \, . 
 \eea

\medskip\noindent
The new connection $\Gamma$ is not invariant under (\ref{gauge-transformation})
but actually transforms like the Weyl gauge field
$\Gamma_{\mu \nu}^\rho \mapsto \Gamma_{\mu \nu}^\rho + \delta_\nu^\rho \partial_\mu \lambda_D$,
and
is not symmetric either, $\Gamma_{\mu\nu}^\rho\not=\Gamma_{\nu\mu}^\rho$; hence
this new formulation is metric but  has torsion!
Then the commutator of the covariant derivatives associated to $\Gamma$ defines
a curvature tensor $R^\rho{}_{\sigma \mu \nu}$ and a torsion depending on $\w_\mu$
\medskip
\begin{equation}\label{t2}
  [\nabla_\mu, \nabla_\nu] V^\rho
  = R^\rho{}_{\sigma \mu \nu} V^\sigma - T_{\mu \nu}{}^\sigma \nabla_\sigma V^\rho \, .
\end{equation}

\medskip\noindent
The tensors defined above can be written explicitly as
\medskip
\begin{align}
R^\rho{}_\sigma{}_{\mu \nu} &= \partial_\mu \Gamma^\rho_{\nu \sigma}
  - \partial_\nu \Gamma^\rho_{\mu \sigma} + \Gamma^\rho_{\mu \tau}
  \Gamma^\tau_{\nu \sigma} - \Gamma^\rho_{\nu \tau} \Gamma^\tau_{\mu \sigma}
  \, , & T_{\mu \nu}{}^\rho = 2 \Gamma_{[\mu \nu]}^\rho = 2 \delta_{[\mu}^\rho \w_{\nu]} \, ,
  \label{curvature-2}
\end{align}

\medskip\noindent
with the usual notation: $A_{[\mu\nu]}\equiv (1/2) (A_{\mu\nu}-A_{\nu\mu})$ (also
$A_{(\mu\nu)}\equiv (1/2) (A_{\mu\nu}+A_{\nu\mu})$).
The trace of the torsion $T_{\mu\nu}{}^{\mu}$ is  proportional to the Weyl gauge field.

The new Weyl-Riemann curvature $R^\rho{}_{\sigma\mu\nu}$ tensor is still invariant 
under dilatations (\ref{gauge-transformation}). This is seen by  using the
projective transformation (\ref{projective}), which relates this formulation
to the previous non-metric one. This gives
\begin{equation}\label{relation}
R^\rho{}_{\sigma \mu \nu} = \tilde R^\rho{}_{\sigma \mu \nu} - \delta^\rho_\sigma F_{\mu \nu}\, ,
\end{equation}
where $F_{\mu\nu}=
\partial_\mu\w_\nu-\partial_\nu\w_\mu$.
Since $F_{\mu\nu}$ and $\tilde R^\rho{}_{\sigma\mu\nu}$ are invariant
under (\ref{gauge-transformation}), then  $R^\rho{}_{\sigma\mu\nu}$
 is also Weyl gauge invariant. The new Weyl-Ricci tensor
$R_{\sigma\nu}=R^\rho{}_{\sigma\rho\nu}$ is also Weyl gauge invariant,
while the  Weyl {geometry} scalar curvature $R=g^{\mu\nu} R_{\mu\nu}$
transforms covariantly, like the inverse metric entering its definition.
Note also that $R=\tilde R$. 

For completeness we present here the Riemannian, Ricci and
scalar curvatures of Weyl geometry {(${R}_{ \rho \sigma \mu \nu }$,  ${R}_{\mu \nu }$, $R$)}
in terms of their Riemannian counterparts
($\mathring{R}_{ \rho \sigma \mu \nu }$,  $\mathring{R}_{\mu \nu }$, $\mathring R$)
obtained by using (\ref{rel}) in (\ref{curvature-2}). A long algebra gives
\begin{align}
    R_{ \rho \sigma \mu \nu } & = \mathring{R}_{ \rho \sigma \mu \nu }
    + \left[g_{\mu \sigma} \mathring{\nabla}_\nu \w_\rho
      - g_{\mu \rho} \mathring{\nabla}_\nu  \w_\sigma
      + g_{\nu \rho}  \mathring{\nabla}_\mu
      \w_\sigma - g_{\nu \sigma}\mathring{\nabla}_\mu \w_\rho  \right] \nonumber\\[2pt]
    &+ \w^2(g_{\mu \sigma} g_{\nu \rho} - g_{\mu \rho} g_{\nu \sigma})
    + \w_\mu(\w_\rho g_{\nu \sigma} - \w_\sigma g_{\nu \rho})
    + \w_\nu(\w_\sigma g_{\mu \rho} - \w_\rho g_{\mu \sigma})\, , \label{riem-1} \\[2pt]
    R_{\mu \nu} & = \mathring{R}_{\mu \nu} - (d-2) \mathring{\nabla}_\nu
    \w_\mu- g_{\mu \nu} \mathring{\nabla}_\rho \w^\rho + (d-2)\w_\mu \w_\nu
    - (d-2)g_{\mu \nu} \w_\rho \w^\rho\, , \label{riem-2}  \\[4pt]
    R &= \mathring{R} - 2(d-1)\, \mathring{\nabla}_\mu \w^\mu
    - (d-1)(d-2)\, \w_\mu \w^\mu \, .  \label{riem-3}
\end{align}

\medskip
Using these results, eq.(\ref{relation}) also gives the relation of
the curvatures in the previous non-metric but torsion-free formulation to their
Riemannian counterparts.

With these observations, a Weyl gauge invariant (quadratic) action can again be build
with these curvature fields. The action so obtained is identical
to that found in the previous formulation, up to a re-definition of its
couplings \cite{CDA}.
However, {similar to} the previous non-metric torsion-free formulation,
the derivatives $\nabla_\alpha$ applied to the curvature tensors,  $\nabla_\alpha
R^\rho{}_{\sigma\mu\nu}$, $\nabla_\alpha R_{\mu \nu}$ and $\nabla_\alpha R$
do not transform covariantly under (\ref{gauge-transformation}).
Hence this {(affine)} formulation is not manifestly Weyl gauge covariant either.

To conclude:
  the previous formulation with {\it vectorial} non-metricity but torsion-free
  was ``traded'' for another formulation that is metric but has {\it vectorial} torsion.
  These two {\it affine formulations}
  are related by a projective transformation eq.(\ref{projective})
  and are equivalent, they lead to the same action, up to a re-definition of the couplings
  \cite{CDA}.
 The explanation for having two equivalent formulations of Weyl conformal geometry
is the following \cite{CDA,CDA2}:
on the tangent space one can define the spin connections $\Om_\mu{}^{ab}$
and $\tilde \Om_\mu{}^{ab}$  to which  $\Gamma$ and
$\tilde \Gamma$ connections are related, via the vielbein postulate. The relation between these
two spin connections can then be seen to correspond to a redefinition of the generators of the
Weyl group (without a change of the  action).
Thus, vectorial torsion and vectorial non-metricity are
not always physical, since neither formulation with these is Weyl gauge covariant;
the formulation guaranteed to be physical is the manifestly  Weyl gauge covariant formulation
which is automatically metric.

\bigskip\noindent
{\bf $\bullet$ Weyl gauge covariant, metric formulation {(non-affine)}}

\bigskip\noindent
We saw that the derivatives $\tilde \nabla_\mu(\tilde\Gamma)$
and $\nabla_\mu(\Gamma)$ of the previous affine formulations
do not transform covariantly with respect to gauged dilatations
transformation (\ref{gauge-transformation})  when acting on the curvature
tensors. Moreover, a  Weyl gauge covariant derivative must be so with
respect to  tensors $X$ of {\it arbitrary} Weyl charge. 
The manifestly Weyl gauge covariant formulation
is based on the space-time gauge covariant derivative $\hat \nabla$ defined below,
which does respect this demand, for $X$ of any Weyl charge. The consequence is that
such formulation is {\it not}  defined by an affine connection anymore, since there is
no $\hat\Gamma$ associated to $\hat\nabla_\mu$ for all $X$ of arbitrary charge.
To understand how this formulation is introduced
recall that in the non-metric formulation  $(\tilde\nabla_\mu+q \,\w_\mu) \,g_{\alpha\beta}=0$
where $q=2$ is the Weyl charge of $g_{\alpha\beta}$. This suggests that for an arbitrary
($r$-contravariant and $p$-covariant) tensor $X$ of arbitrary charge,  define $\hat\nabla$:
\medskip
\begin{equation}
\label{general-formula}
\qquad\hat \nabla_\mu X
\equiv \Big[\tilde \nabla_\mu(\tilde\Gamma)  + \tilde q_{X} \,\w_\mu\Big]\, X
= \Big[\nabla_\mu(\Gamma)  + q_{X} \,\w_\mu\Big]\, X \, ,\qquad
\textrm{where} \quad X\equiv X^{\mu_1 \ldots \mu_r}_{\nu_1 \ldots \nu_p}\, ,
\end{equation}
with $\tilde q_X$ the space-time Weyl charge. In the second step
we wrote explicitly the action of $\tilde\nabla_\mu$ on $X$ and
then  replaced $\tilde\Gamma$ by $\Gamma$ eq.(\ref{projective});
hence  $q_X$ denotes the tangent space-time Weyl charge of $X$.
In this way we find the two charges are related by
\begin{equation}
  \label{charge}
 q_{X} = \tilde q_{X}+r-p  \, \qquad\textrm{where}\qquad X\equiv X^{\mu_1...\mu_r}_{\nu_1...\nu_p}\, .
\end{equation}

One can work with either the first or the second definition of $\hat\nabla_\mu$ in
(\ref{general-formula}), since they are equivalent.
The first is using space-time charge $\tilde q$, the second is using the corresponding
tangent space charge $q$ of the same  $X$. Most of the results in this paper were derived in both approaches (using space-time or tangent space charges).
To see one difference, note that a tangent space vector $V^a$ that is invariant
under (\ref{gauge-transformation})  has $q=0$, then the space-time charge
$\tilde q(V^\mu)=\tilde q(V^a e_a^\mu)=-1$ since $g^{\mu\nu}$ has space-time
charge $\tilde q_{g^{\mu\nu}}=-2$ so  $\tilde q(e_a^\mu)=-1$. However
$\tilde q(V_\mu)=\tilde q(g_{\mu\nu} V^\nu)=+1\not=\tilde q(V^\mu)$; hence
attention must be paid to the space-time charge of each covariant  or
contravariant degree of an object when working  with $\tilde\nabla_\mu$.
If one is using $\nabla_\mu$ instead, the tangent space Weyl charge is
independent of this aspect, since the Minkowski metric $\eta_{ab}$ has $q=0$.

In the table below we give the space-time charges $\tilde q_X$ and tangent space
charges $q_X$ of various quantities $X$ which are used  in expressing the
action of  the gauge covariant derivative $\hat \nabla$ in terms of
$\tilde \nabla$ or $\nabla$, according to \eqref{general-formula}.
\begin{table}[ht]
  \label{charges}
  \centering
  \begin{tabular}{c|ccccccccccc} 
 $X$ & $e_\mu^a$ & $e_a^\mu$ & $\eta_{ab}$ & $g_{\mu\nu}$ & $g^{\mu\nu}$ &  $\sqrt{g}$
    &  $ \sqrt{g}\,\epsilon_{\mu_1...\mu_d}$ & $R^{\rho}{}_\sigma{}_{\mu\nu}$ & $R_{\mu\nu}$ & $R$ & $F_{\mu\nu}$
    \\[2mm]
    \hline
 $\tilde q_X$ &   1 & -1 & 0 & 2 & -2 & $d$ &   $d$ & 0& 0 & -2 & 0
    \\[2mm]
        $ q_X$ &   0 & 0 & 0 & 0 & 0  & $d$ &   $0$ & -2& -2 & -2 & -2
    \\[2mm]
  \end{tabular} \, .
\end{table}
Strictly speaking tangent space charges  or space-time charges for the
vielbein ($e_\mu^a$) do not make sense as this is a ``mixed'' object (with both types of indices).
In the table,
the dilatation charge of $e_\mu^a$  is actually associated to the curved space-time
  index on which the space-time derivative $\hat \nabla$ acts
  (see \eqref{viel-1}-\eqref{viel-2}).

The Weyl gauge-covariant derivative $\hat \nabla$ ensures  that  for
an arbitrary tensor $X$ (of charge $\tilde q_X$ (equivalently $q_X$)), 
that transforms in a Weyl gauge covariant way under (\ref{gauge-transformation}),
its derivative $\hat \nabla_\mu X$ also transforms in a similar way,
with the same charge\footnote{
As seen from  \eqref{general-formula}, the affine  derivatives
$\tilde \nabla$ (or $\nabla$) can act in a gauged dilatation covariant way {\it only} on
tensors with vanishing  space-time Weyl charge (vanishing tangent space Weyl charge),
respectively.} 
\medskip
\begin{align}
  X \mapsto e^{\tilde q_X \lambda_D} X\, , &&
  \hat \nabla_\mu X \mapsto e^{\tilde q_{X} \lambda_D} \hat \nabla_\mu X \, ,
\label{gen-transf}
\end{align} 
where we did not display the tensor indices of $X$.
The Weyl gauge covariant derivative $\hat\nabla$ is compatible both with
the metric and the completely antisymmetric tensor
\begin{align}
  \hat \nabla_\mu g_{\nu \rho} & = 0\, , & \hat \nabla_\mu
  \left( \sqrt{g} \, \epsilon_{\nu \rho \sigma \delta} \right) & = 0 \, .
\label{metric-theory}
\end{align}

\medskip\noindent
Therefore, in this formulation  the theory is indeed metric and manifestly gauge covariant, something
that is obscured in the historical non-metric
formulation (based on $\tilde\nabla(\tilde\Gamma)$) or in that with
torsion (based on $\nabla(\Gamma)$). As a result of
Weyl gauge covariance and metricity, the parallel transport
of a (Weyl gauge covariant) vector automatically preserves the norm of this
vector - a natural result (in a gauge theory gauge covariance
must be respected by the parallel transport, in order for this transport
to be physically meaningful).

On an arbitrary vector $V^\rho$ the commutator of covariant derivatives can be expressed as
\medskip
\begin{equation}
  \label{hatcomm}
  \big [ \hat \nabla_\mu, \hat \nabla_\nu \big ] V^\rho 
  = \tilde R^\rho{}_{\sigma \mu \nu}
  V^\sigma+ \tilde q_V F_{\mu \nu} V^\rho
  =  R^\rho{}_{\sigma \mu \nu}  V^\sigma+ q_V F_{\mu \nu} V^\rho \, ,
\end{equation}

\medskip\noindent
where $R^\rho{}_{\sigma \mu \nu}$, $\tilde R^\rho{}_{\sigma \mu \nu}$ and $F_{\mu \nu}$
given in eqs.\,\eqref{curvature-2},\eqref{curvature-1},\eqref{curvature-3}
are regarded here as gauge curvatures. 
Again, we expressed the lhs either in terms of space-time charge
or tangent space charge. We used either (\ref{t1}) or (\ref{t2}) and then
(\ref{relation}), (\ref{charge}).

The transformations of the space-time quantities introduced thus far and of
their  gauge covariant derivatives are found from the general
formula \eqref{gen-transf}
\medskip
\begin{align}
  g_{\mu \nu} & \mapsto e^{2 \lambda_D} g_{\mu \nu}\, , & F_{\mu \nu} &\mapsto F_{\mu \nu}
  \, , & \hat \nabla_\alpha F_{\mu \nu} &\mapsto \hat \nabla_\alpha F_{\mu \nu}\, ,\nonumber\\
  R^\rho{}_{\sigma \mu \nu} &\mapsto R^\rho{}_{\sigma \mu \nu}\, , & R_{\mu \nu} &\mapsto R_{\mu \nu}\, ,
  & R &\mapsto e^{-2 \lambda_D} R\, , \label{cov1}\\
  \hat \nabla_\alpha R^\rho{}_{\sigma \mu \nu} &\mapsto \hat \nabla_\alpha R^\rho{}_{\sigma \mu \nu}\, ,
  & \hat \nabla_\alpha R_{\mu \nu} &\mapsto \hat \nabla_\alpha R_{\mu \nu}\, ,
  & \hat \nabla_\alpha R &\mapsto e^{-2 \lambda_D} \hat \nabla_\alpha R\, , \nonumber
\end{align}
and also
\begin{align}
R_{\mu\nu\rho\sigma}^2 & \mapsto  e^{-4\lambda_D} \, R_{\mu\nu\rho\sigma}^2\, , &
R_{\mu\nu}^2 & \mapsto e^{-4\lambda_D}\,R_{\mu\nu}^2\, , &
R^2 &\mapsto e^{-4\lambda_D}\,R^2\, ,
\nonumber \\
F_{\mu\nu}^2 & \mapsto  e^{-4\lambda_D}\,F_{\mu\nu}^2\, , &
G & \mapsto e^{-4\lambda_D} G\, , &
C_{\mu \nu \rho \sigma}^2 &\mapsto e^{-4\lambda_D}  C_{\mu \nu \rho \sigma}^2\, .\label{cov2}
\end{align}
$G$ is the  expression of the Chern-Euler-Gauss-Bonnet term  (hereafter Euler term)
valid in the presence of torsion 
\bea\label{G}
G = R_{\mu \nu \rho \sigma} R^{\rho \sigma \mu \nu} - 4 R_{\mu \nu}R^{\nu \mu} + R^2 \, .
\eea

\medskip\noindent
$C_{\mu\nu\rho\sigma}$ is the Weyl tensor of Weyl geometry in the covariant
formulation eq.(\ref{D6}), in which it is identical to its
Riemannian geometry counterpart, see eq.(\ref{Cw}):
$C_{\mu \nu \rho \sigma} = \mathring C_{\mu \nu \rho \sigma}$. That $G$ transforms
covariantly, see (\ref{cov2}),  is an important feature specific to conformal geometry.
Formulae  identical to (\ref{cov1}), (\ref{cov2}) also
apply for $\tilde R^\rho{}_{\sigma \mu \nu}$, $\tilde R_{\nu\sigma}$, $\tilde R$
and their squares\footnote{
Note that the other differential operators $\nabla_\mu R$
or $\tilde \nabla_\mu \tilde R$ do not transform covariantly.}.

A practical aspect of the gauge covariant formalism is the
integration by parts when using the gauge covariant derivative $\hat\nabla$.
It will be used later when deriving the equations of motion. With
the formulae already given, one can show that (in $d$ dimensions)
\medskip
\begin{equation}\label{p}
  \int d^d x\, \sqrt{g}\,\,  \hat \nabla_\mu J^\mu  = \int d^d x \,
  \partial_\mu \,\left(\sqrt{g} J^\mu \right) + (d+\tilde q_J)
  \int d^dx \, \sqrt{g}\,\, \w_\mu J^\mu\, ,
\end{equation}
where we used (\ref{general-formula}) and
 $\tilde\nabla_\mu J^\nu=\mathring\nabla_\mu J^\nu +(\delta_\mu^\nu\w_\lambda
+\delta_\lambda^\nu \w_\mu - g_{\mu\lambda}\w^\nu)\,J^\lambda$, so $\tilde\nabla_\mu J^\mu=
\mathring\nabla_\mu J^\mu +d\,\w_\lambda\,J^\lambda$ and
$\sqrt{g}\mathring\nabla_\mu J^\mu=\partial_\mu(\sqrt{g}\,J^\mu)$.
If  the integral  of the lhs is  invariant under gauged
dilatations (\ref{gauge-transformation})
i.e.  the charge $\tilde q_{J^\mu}=-d$, then the last term
in the rhs in (\ref{p}) vanishes. Then the
result is similar to that in Riemannian geometry, with $\hat\nabla$
replaced by $\mathring\nabla$.

Next, let us define a vector $J^\mu \equiv X^{\mu \nu_1 \ldots \nu_p} Y_{\nu_1 \ldots \nu_p}$, 
with tensors $X$ and $Y$ both covariant under (\ref{gauge-transformation}) and with
$\tilde q_{J^\mu} \equiv  \tilde q_{X^{\mu \nu_1...\nu_p}} + \tilde q_{Y_{\nu_1...\nu_p}}$.
Assume that $\tilde q_{J^\mu}+d=0$ which means the lhs of the relation
below is Weyl gauge invariant - a common situation encountered in this work;
in such case 
\medskip
\begin{equation}
  \int d^d x \, \sqrt{g}\, X^{\mu \nu_1 \ldots \nu_p} \hat \nabla_\mu Y_{\nu_1 \ldots \nu_p} =
  \int d^d x \,  \partial_\mu \left(\sqrt{g} J^\mu \right)
  - \int d^dx \, \sqrt{g} \, \left(\hat \nabla_\mu X^{\mu \nu_1 \ldots \nu_p} \right)
  Y_{\nu_1 \ldots \nu_p} \, . 
\label{parts}
\end{equation}

\medskip\noindent
Relations (\ref{p}), (\ref{parts}) are  useful in practice, when deriving
the equations of motion, etc. For  similar results  in the Weyl gauge covariant
metric formulation, see the Appendix.

Therefore, for actions invariant under gauged dilatations (and for similar behaviour
at infinity/boundary) one can integrate by parts  just like in the Riemannian case;
in this sense, $\hat\nabla$ is for  Weyl geometry  what $\mathring\nabla$ is for
Riemannian geometry. This clarifies the action of Weyl gauge covariant derivative
$\hat\nabla$ - it enables calculations like in Riemannian geometry.

To complete the geometric part of the gauge covariant formalism, we present
here the Bianchi identities written with the help of $\hat \nabla$
\medskip
\begin{align}
  R^\sigma{}_{[\mu \nu \rho]} & = - \delta_{[\mu}^\sigma F_{\nu \rho]}\, , &
  \hat \nabla_{[\mu} R^{\tau \sigma}{}_{\nu \rho]} & = 0\, , & \hat \nabla_{[\mu} F_{\nu \rho]} = 0 \, .
\label{bianchi-covariant}
\end{align}

\medskip\noindent
Additional  useful identities for the gauge covariant formalism 
are shown  in the Appendix.

We can now  present the  most general action of  Weyl gravity
associated to conformal geometry, regarded as a gauge theory of the
Weyl group,  in the Weyl gauge covariant formalism.
The analysis so far is valid in $d$ dimensions, so we  present
this action for $d=4-2\epsilon$ \cite{DG1,CDA} (see also \cite{WDBI})
\medskip
\begin{equation}
  S = \int d^d x \sqrt{g}\, \left[\alpha_1 R^2
    + \alpha_2 C_{\mu \nu \rho \sigma} C^{\mu \nu \rho \sigma}
    + \alpha_3 F_{\mu \nu} F^{\mu \nu} + \alpha _4 G \right]R^{(d-4)/2} \,. 
\label{action-d}
\end{equation}

\medskip\noindent
Each term in this  action is Weyl gauge invariant, see (\ref{cov2}).
Note that the  Euler term $G$  contributes to the equations of motion
if $d\neq 4$. As discussed elsewhere \cite{CDA} up to a redefinition of
the  {\it dimensionless} couplings $\alpha_i$, this action has the same
form in the previous two formulations (non-metric torsion-free
and metric formulation but with torsion).

Note that action (\ref{action-d})  can be expressed in
  a Riemannian notation using eq.(\ref{riem-3}) for $ R$, eq.(\ref{G}) with
  replacements (\ref{riem-1}) to (\ref{riem-3}) for $G$,
  while the Weyl tensor in this Weyl covariant formulation is the same in both
  geometries, $C_{\mu\nu\rho\sigma}=  \mathring{C}_{\mu\nu\rho\sigma}$, as mentioned.

In summary, the Weyl gauge covariant formalism is defined by $\hat \nabla$ which
can be expanded \eqref{general-formula} either in terms of $\Gamma$ or
$\tilde \Gamma$ for intermediate steps of calculation, but in general differs
from both of them for non-zero charges. With $\hat\nabla$, Weyl geometry appears as a
``covariantised'' version of Riemannian geometry with respect to gauged dilatations.
We have two gauge curvatures: $R^\rho{}_{\sigma \mu \nu}$ associated
to the Lorentz subgroup and $F_{\mu\nu}$ associated to the dilatations which
arise on the right-hand-side of the commutator of gauge covariant
derivatives. These curvatures and their $\hat \nabla$ derivatives
transform covariantly under dilatations \eqref{cov1}. The theory is metric
as seen from \eqref{metric-theory}. Regarding the previous two formulations:
non-metric torsion-free vs metric with torsion, passing from one to the
other  amounts to a projective transformation \eqref{projective}
which in turn corresponds to a redefinition of the generators of the Weyl
group. These formulations can be used, but they do not always lead to
physical results, unlike the Weyl gauge covariant formulation
that keeps manifest Weyl gauge symmetry.

The formulae for integration by parts with the gauge covariant derivative
\eqref{parts} are just like in the Riemannian case, so in this sense $\hat\nabla$ is
for Weyl geometry what $\mathring\nabla$ is for Riemannian geometry.
The Bianchi identities can be written in manifestly covariant
form \eqref{bianchi-covariant}, similar in structure to their Riemannian version.
For an arbitrary dimension $d=4-2\epsilon$, relevant for defining the theory
at the quantum level, the Lagrangian remains Weyl gauge invariant
due to a ``geometric'' regularisation with $ R^{(d-4)/2}$ - a feature
specific to Weyl geometry that maintains manifest Weyl gauge invariance
of the action, while its couplings remain dimensionless.

\section{Equations of motion with Weyl gauge covariance}\label{3}

In this section we derive  the equations of motion of the most general Weyl gravity action
in four dimensions, in a manifestly gauged dilatation covariant form.

For $d=4$,  action \eqref{action-d} {in Weyl geometry}  becomes 
\begin{equation}
  S = \int d^4 x \sqrt{g}\, \left[\alpha_1 R^2
    + \alpha_2 C_{\mu \nu \rho \sigma} C^{\mu \nu \rho \sigma}
    + \alpha_3 F_{\mu \nu} F^{\mu \nu} + \alpha _4 G \right]\, .
\label{action}
\end{equation}

\medskip\noindent
To keep notation simple, we use  dimensionless coefficients $\alpha_i$, $i=1,2,3,4$,
but the actual physical perturbative couplings, denoted 
$\xi,\eta,\alpha$ are shown below \cite{SMW}
\bea
\alpha_1=\frac{1}{\xi^2}\, , \qquad
\alpha_2= -\frac{1}{\eta^2}\, , \qquad
\alpha_3=-\frac{1}{4\alpha^2}\, , \qquad
\alpha_4=1,\qquad \xi,\eta,\alpha<1\, .
\eea

\medskip
In four dimensions the Euler term
$G$ does not contribute to the equations of motion.
To derive these equations  it is useful to introduce the
following basis of integrals
\begin{equation}\label{eq}
\begin{split}
I_1 &= \int d^4x \, \sqrt{g}\, R^2\, , \\
I_2 & = \int d^4 x \, \sqrt{g}\, R_{\mu \nu}R^{\mu \nu} \, ,
\end{split}
\qquad \qquad
\begin{split}
  I_3 &= \int d^4x \, \sqrt{g}\, F_{\mu \nu} F^{\mu \nu}\, , \\
   I_4 & = \int d^4 x \, \sqrt{g}\, R_{\mu \nu \rho \sigma} R^{\mu \nu \rho \sigma} \, .
\end{split}
\end{equation}

\medskip\noindent
Action \eqref{action} can be written as
\begin{equation}\label{eq32}
  S = a_1 I_1 +a_2 I_2 +a_3 I_3 + a_4 I_4 \, ,
\end{equation}
where
\begin{align}
  a_1   = \alpha_1 + \frac{\alpha_2}{3} + \alpha_4 \, , \quad
  a_2  = - (2 \alpha_2 + 4 \alpha_4) \, ,\quad
  a_3  = \alpha_3 + 4 \alpha_4\, ,\quad
  a_4  = \alpha_2 + \alpha_4 \, .\quad
\label{a-alpha}
\end{align}
These  values  of  $a_i$ arise from writing  the Weyl tensor term  and the Euler
term in the action as a  function of integrands  (\ref{eq}),
using (\ref{weyl-squared}), (\ref{D9}) in Appendix~D.

The variation  $\delta S$ of the action is then expressed in terms of the variations
$\delta_g I_i$ with respect to the metric and $\delta_\w I_i$ with respect to $\w_\mu$, where
\begin{align}\label{def-WB}
  \delta_g I_i &= \int d^4 x\, \sqrt{g}\, W_{\mu \nu}^{(i)}\,
  \delta g^{\mu \nu}\, , & \delta_\w I_{i} & =  \int d^4 x\,
  \sqrt{g}\, B^{(i)\, \mu}\, \delta \w_\mu \, , \qquad i=1,2,3,4\, .
\end{align}

\medskip\noindent
with symmetric tensors $W_{\mu \nu}^{(i)} = W_{(\mu \nu)}^{(i)}$ and
$B_\mu^{(i)}$ to be computed below. 
Further, we denote
\medskip
\bea\label{eqs2}
W_{\mu\nu}\equiv \frac{1}{\sqrt{g}}
\frac{\delta S}{\delta g^{\mu\nu}}=\sum_{i=1}^4 a_i \,W_{\mu\nu}^{(i)}\, ,\qquad\quad
\textrm{and}\qquad\quad
B_\mu\equiv \frac{g_{\mu\nu}}{\sqrt{g}} 
\frac{\delta S}{\delta \w_\nu}=\sum_{i=1}^4 a_i \,B_\mu^{(i)}\, .
\eea

\medskip\noindent
Therefore $\delta S=\int d^4x\sqrt{g}\, (W_{\mu\nu} \delta g^{\mu\nu}+ B^\mu\,\delta \w_\mu)$
so the equations of motion of $g_{\mu\nu}$ and $\w_\mu$ are
\medskip
\bea\label{eqs}
W_{\mu\nu}=0\, ,\qquad\qquad B_\mu=0 \, .
\eea

\medskip\noindent
If a given term in the general action (\ref{action}) is absent, one restricts the
sums in (\ref{eqs2}) accordingly. 

Let us compute the expressions of $W_{\mu\nu}^{(i)}$ and  $B_\mu^{(i)}$.
For this, we  first calculate  the variation $\delta R^\rho{}_{\sigma\mu\nu}$ of
{the  Weyl-Riemann tensor} $R^\rho{}_{\sigma \mu \nu}$ under an arbitrary
variation $\delta \Gamma_{\mu\nu}^\rho$ of $\Gamma$:
\medskip
\begin{equation}
  \label{deltaR0}
  \delta R^\rho{}_{\sigma\mu\nu}=
  R^\rho{}_{\sigma\mu\nu}(\Gamma+\delta\Gamma)-R^\rho{}_{\sigma\mu\nu}(\Gamma) \, .
\end{equation}

\medskip\noindent
In this formula we use the first equation
in (\ref{curvature-2}) and then Taylor expand the result to linear order in $\delta\Gamma$.
Next, we compute the difference
$\nabla_\mu\delta\Gamma_{\nu\sigma}^\rho-\nabla_\nu\delta\Gamma_{\mu\sigma}^\rho$,
by writing the standard action of $\nabla(\Gamma)$ on
$\delta\Gamma_{\nu\sigma}^\rho$ and $\delta\Gamma_{\mu\sigma}^\rho$,
then  use this difference in the above formula for
$\delta R^\rho{}_{\sigma\mu\nu}$. In these steps,  we use that the variation
$\delta \Gamma$ is a tensor (even though $\Gamma$ itself is not!),
and that $\delta\Gamma=\Gamma-\Gamma'$ is invariant under Weyl
gauge transformations (i.e. $\delta \Gamma$ has a vanishing space-time charge
$\tilde q_{\delta\Gamma}=0$ or tangent space charge  $q_{\delta\Gamma} = -1$). This invariance
follows since both $\Gamma$ and $\Gamma'$ transform like the Weyl gauge boson, as
mentioned. We obtain
\begin{equation}
  \label{eqR}
  \delta R^\rho{}_{\sigma\mu\nu}=
  \nabla_\mu \delta\Gamma_{\nu\sigma}^\rho-\nabla_\nu\delta\Gamma_{\mu\sigma}^\rho
  +2 \Gamma^\lambda_{[\mu\nu]}\,\delta \Gamma_{\lambda\sigma}^\rho \, .
\end{equation}

\medskip\noindent
Further,  use the second relation in \eqref{curvature-2},
$2 \Gamma^\lambda_{[\mu\nu]}=\delta_\mu^\lambda \w_\nu-\delta_\nu^\lambda\w_\mu$, 
and that the tangent space charge of $\delta \Gamma$ is $q_{\delta\Gamma} = -1$, then with
(\ref{general-formula})  the above formula becomes\footnote{
A similar formula can be written for $\delta\tilde R^\rho{}_{\sigma \mu \nu}$.}
\medskip
\begin{equation}
  \label{deltaRiem}
  \delta R^\rho{}_{\sigma \mu \nu} =
  \hat \nabla_\mu \delta\Gamma_{\nu \sigma}^\rho
  - \hat \nabla_\nu \delta \Gamma_{\mu \sigma}^\rho \, .
\end{equation}
It then follows that
\medskip
\begin{align}\label{deltaR}
  \delta R_{\mu \nu} &= \hat \nabla_\rho \delta \Gamma_{\nu \mu}^\rho
  - \hat \nabla_\nu \delta \Gamma_{\rho \mu}^\rho\, , & \delta R
  = \big(\hat \nabla_\rho \delta \Gamma_{\nu \mu}^\rho
  - \hat \nabla_\nu \delta \Gamma_{\rho \mu}^\rho \big)\, g^{\mu \nu} + R_{\mu \nu} \delta g^{\mu \nu} \, .
\end{align}

So far $\delta \Gamma$ was arbitrary. Since $\Gamma$ is not an independent field in
the theory but is given by \eqref{rel}, the variation  $\delta\Gamma$ can be due to
a variation  of the metric ($\delta g_{\mu\nu}$), of $\w_\mu$ ($\delta \w_\mu$)
or of both. For the Levi-Civita connection $\mathring \Gamma$ we have the well-known result
for its variation $\delta_g\mathring\Gamma$  due to  the metric variation $\delta g_{\mu\nu}$
\begin{equation}
  \delta_g \mathring \Gamma_{\mu \nu}^\rho =\frac12 g^{\rho \sigma}
  \big(\mathring \nabla_\mu \delta g_{\nu \sigma}
  + \mathring \nabla_\nu \delta g_{\mu \sigma} - \mathring \nabla_\sigma \delta g_{\mu \nu} \big) \, .
\end{equation}

\medskip\noindent
Similarly, for the variation $\delta_g \Gamma$ of $\Gamma$
due to  a variation $\delta g_{\mu\nu}$ for $\w_\mu$ {\it fixed},
we find from (\ref{rel})\footnote{Equivalently one can work with $\tilde \Gamma$.}
\medskip
\begin{equation}\label{eqD}
  \delta_g \Gamma_{\mu \nu}^\rho = \frac12 g^{\rho \sigma}
  \left(\nabla_\mu \delta g_{\nu \sigma} +
  \nabla_\nu \delta g_{\mu \sigma} - \nabla_\sigma \delta g_{\mu \nu} \right) \, .
\end{equation}

\medskip\noindent
Next, the variation $\delta g_{\mu\nu}$ has the same Weyl charge as the metric,
which is actually vanishing on the tangent space, therefore
according to (\ref{general-formula}),  in the formula above
the action of $\hat\nabla$ is the same as that of $\nabla$; 
 we thus obtain a  manifestly gauged  dilatation covariant form 
\medskip
\begin{equation}
  \delta_g \Gamma_{\mu \nu}^\rho = \frac12 g^{\rho \sigma}
  \big(\hat \nabla_\mu \delta g_{\nu \sigma}
  + \hat \nabla_\nu \delta g_{\mu \sigma} - \hat \nabla_\sigma \delta g_{\mu \nu} \big)\, .
\end{equation}

\medskip\noindent
We obtain exactly the same result for the variation $\delta_g\tilde\Gamma_{\mu\nu}^\rho$.
This equation also confirms that the space-time Weyl charge 
of $\delta \Gamma$ is vanishing $\tilde q_{\delta\Gamma}=0$.
Using this equation in (\ref{deltaR}), we finally find the variations of
$R^\rho{}_{\sigma \mu \nu}$, $R_{\mu \nu}$ and $R$ induced by the metric variation
\begin{align}
  \delta_g R^\rho{}_{\sigma \mu \nu} & = \frac12 g^{\rho \lambda}
  \left(\hat \nabla_\mu \hat \nabla_\nu \delta g_{\sigma \lambda}
  + \hat \nabla_\mu \hat \nabla_\sigma \delta g_{\nu \lambda}
  - \hat \nabla_\mu \hat \nabla_\lambda \delta g_{\nu \sigma}
  - \hat \nabla_\nu \hat \nabla_\mu \delta g_{\sigma \lambda} \right. \nonumber\\
  & \left. - \hat \nabla_\nu \hat \nabla_\sigma \delta g_{\mu \lambda}
  + \hat \nabla_\nu \hat \nabla_\lambda \delta g_{\mu \sigma}\right)\, , \label{var-1}
  \\
  \delta_g R_{\mu \nu} & = \frac12 \,\big(\, g_{\mu \rho} g_{\nu \sigma} \hat \Box
  + g_{\rho \sigma} \hat \nabla_\nu \hat \nabla_\mu
  - g_{\mu \rho} \hat \nabla_\sigma \hat \nabla_\nu
  - g_{\nu \rho} \hat \nabla_\sigma \hat \nabla_\mu\,\big) \,\delta g^{\rho \sigma}\, ,
  \label{var-2}
  \\[4pt]
  \delta_g R & = \big(-\hat \nabla_\mu \hat \nabla_\nu
  + R_{\mu \nu} + g_{\mu \nu} \hat \Box \,\big)\, \delta g^{\mu \nu} \, . \label{var-3}
\end{align}  

\medskip\noindent
To go from  $\delta g_{\mu\nu}$ to $\delta g^{\mu\nu}$ one  uses
that  $g^{\mu \rho} \delta g_{\rho \nu} = - g_{\rho \nu} \delta g^{\mu \rho}$
derived from  $\delta (g_{\mu\nu} g^{\nu\sigma})=0$.

Notice that in this  Weyl gauge covariant formulation,
the variations in eqs.(\ref{var-1}) to (\ref{var-3}) have the same form as those in
the Riemannian geometry in its corresponding notation.

Consider next the variation $\delta_\w\Gamma$ due to
an arbitrary variation of the Weyl gauge field  $\delta \w_\mu$
while keeping the metric fixed. Under such variation, $\Gamma$ changes as, \eqref{rel} 
\medskip
\begin{equation}
  \label{var-4}
  \delta_\w \Gamma_{\mu \nu}^\rho = \left(\delta_\mu^\rho \delta_\nu^\sigma
  - g_{\mu \nu} g^{\rho \sigma} \right) \delta \w_\sigma\, , 
\end{equation}
which is again Weyl invariant as argued before.
Inserting this in \eqref{deltaRiem} and \eqref{deltaR} we obtain
\begin{align}
  \delta_\w R_{\rho \sigma \mu \nu} & = g_{\mu \sigma} \hat \nabla_\nu \delta \w_\rho
  - g_{\mu \rho} \hat \nabla_\nu \delta \w_\sigma
  + g_{\nu \rho} \hat \nabla_\mu \delta \w_\sigma
  - g_{\nu \sigma} \hat \nabla_\mu \delta \w_\rho\, , \label{var-5}\\
  \delta_\w R_{\mu \nu} & = -2 \hat \nabla_\nu \delta \w_\mu
  - g_{\mu \nu} \hat \nabla^\rho \delta \w_\rho\, , \label{var-6}\\
\delta_\w R & = -6 \hat \nabla^\rho \delta \w_\rho\, , \label{var-8}
\end{align}
while for the field strength $F$ we have
\begin{equation}
  \label{var-7}
  \delta_\w F_{\mu \nu} = 2 \hat \nabla_{[\mu} \delta \w_{\nu]}\, .
\end{equation}
As a side-remark, note that the results in eqs.(\ref{deltaR0}) to (\ref{var-7}) are actually
valid in $d$ dimensions.

With these  variations, we  find the expressions of the tensors $W_{\mu \nu}^{(i)}$ of
eq.(\ref{eqs2}) in $d=4$
\medskip
\begin{align}
  W_{\mu \nu}^{(1)} &= 2 g_{\mu \nu} \hat \Box R - 2\hat \nabla_\nu \hat \nabla_\mu R
  + 2 R_{\mu \nu} R - \frac12 g_{\mu \nu} R^2\, , \label{W1}\\
  W_{\mu \nu}^{(2)} & =  g_{\mu \nu} \hat \nabla_\rho \hat \nabla_\sigma R^{\rho \sigma}
  + \hat \Box R_{(\mu \nu)} - \hat \nabla_\sigma \hat \nabla_{(\nu} g_{\mu) \rho} R^{\rho \sigma}
  - \hat \nabla_\rho \hat \nabla_{(\nu} g_{\mu) \sigma} R^{\rho \sigma}\nonumber\\
  &+ (R_{\mu \rho} R_{\nu \sigma} + R_{\rho \mu} R_{\sigma \nu})g^{\rho \sigma}
  - \frac12 g_{\mu \nu} R_{\rho \sigma} R^{\rho \sigma}\, ,\label{W2}\\
  W_{\mu \nu}^{(3)} & = 2 F_{\mu \rho} F_\nu{}^\rho
  - \frac12 g_{\mu \nu} F_{\rho \sigma} F^{\rho \sigma}\, , \label{W3}\\
  W_{\mu \nu}^{(4)} & = - 4 \hat \nabla_\sigma \hat \nabla_\rho  R^\sigma{}_{(\mu \nu)}{}^ \rho
  + 2  R_{\rho \sigma \alpha \mu} R^{\rho \sigma \alpha}{}_\nu
  - \frac12 g_{\mu \nu}  R_{\rho \sigma \alpha \beta} R^{\rho \sigma \alpha \beta}\, . \label{W4}
\end{align}

\medskip\noindent
Note that the first equation of motion is symmetric due to the
identity $ -2 \hat \nabla_{[\nu} \hat \nabla_{\mu]} R + 2 R_{[\mu \nu]} R = 0 $
which follows from the commutator of gauge covariant derivatives
\eqref{hatcomm} particularised for the scalar $R$ 
and a contraction with the metric of the first Bianchi identity.
In eq.\,\eqref{W2} one can further use the identity
$ g_{\mu \nu} \hat \nabla_\rho \hat \nabla_\sigma R^{\rho \sigma}
=\frac12 g_{\mu \nu} \hat \Box R$ (valid only in $d=4$). 

Finally, we are considering the variations with respect to the
Weyl gauge field $\w_\mu$. In this case from eqs.\,\eqref{var-4}-\eqref{var-8} one obtains
in $d=4$ dimensions
\medskip
\begin{align}
B_\mu^{(1)} & = 12 \hat \nabla_\mu R\, , \label{B1}\\
B_\mu^{(2)} & = 4 \hat \nabla^\nu R_{\mu \nu} + 2 \hat \nabla_\mu R
= 4 \hat \nabla_\mu R + 8 \hat \nabla_\nu F_\mu{}^\nu\, , \label{B2}\\
B_\mu^{(3)} & =  - 4 \hat \nabla_\nu F^\nu{}_\mu\, , \label{B3}\\
B_\mu^{(4)} & = 8 \hat \nabla^\nu R_{\mu \nu}   = 4 \hat \nabla_\mu R +
16 \hat \nabla^\nu F_{\mu\nu}\, , \label{B4}
\end{align}

\medskip\noindent
where for the second form of $B_\mu^{(2)}$ and $B_\mu^{(4)}$ we used
the identities $\hat \nabla_\mu R^\mu{}_\nu = \frac12 \hat \nabla_\nu R$, $R_{\mu \nu}
= R_{\nu \mu} + (d-2) F_{\mu \nu}$ with $d=4$ following from the Bianchi
identities \eqref{bianchi-covariant}. 

The general equations of motion of Weyl quadratic gravity in $d=4$
are then
%
\begin{align}\label{eom-1}
  c_1\, W_{\mu \nu}^{(1)}\, +
  c_2\, W_{\mu \nu}^{(2)} +
  c_3 \, W_{\mu \nu}^{(3)} & = 0 \, ,\\[5pt]
  c_1\, B_\mu^{(1)}\,\, +
  c_2 \, B_\mu^{(2)}+
  c_3\, B_\mu^{(3)} & =0\, ,\qquad\label{eom-2}
\end{align}
%
where we eliminated $W_{\mu \nu}^{(4)}$ and $B_\mu^{(4)}$ by 
using that the Euler term $G$ does not contribute to the equations
of motion in $d=4$.
The  coefficients $c_i$ are  given by
\begin{align}
c_1&=a_1-a_4=\alpha_1-(2/3) \,\alpha_2=1/\xi^2+2/(3\eta^2)\, ,\\
c_2&=a_2+4 \,a_4=2\,\alpha_2=-2/\eta^2\, ,\\
c_3&=a_3-4 \, a_4=\alpha_3-4\,\alpha_2=-1/(4\alpha^2)+4/\eta^2\, .
\end{align}
%
The expressions  of the coefficients $c_i$ 
detail which terms in the action and  variations
(with respect to the metric and $\w_\mu$)  actually  contribute to
the equations of motion. If one or more  terms in the general action
are absent, the equations of motion simplify in an obvious way, by setting
to zero the corresponding couplings.

An important result  is that all these equations of motion, derived
from the most general action of gravity (as a gauge theory), are
manifestly covariant with respect to gauged dilatations,
as  actually expected in  a gauge theory. This is because $W_{\mu\nu}^{(i)}$ and $B_\mu^{(i)}$
transform covariantly. This result  validates  our Weyl gauge  covariant formalism.
These equations can be used in applications, in the conformal geometry
picture. They can also be ``translated'' into the Riemannian picture
by   expressing the curvature terms in (\ref{eom-1}), (\ref{eom-2})
in terms of their Riemannian counterparts.
We shall study  these equations for particular cases in section~\ref{5}.

\section{Conservation laws in Weyl conformal geometry in $d=4$}

\subsection{Conservation laws} 

We can now find the conservation laws for
the general Weyl quadratic gravity action, eq.(\ref{action}).

In general relativity the Einstein tensor is covariantly conserved.
Since $W^{(i)}_{\mu \nu}$ are the generalisation of the Einstein tensor to Weyl
quadratic gravity, we expect similar conservation. However, since the
tensors in eqs. \eqref{W1} to \eqref{W4} are manifestly Weyl gauge
covariant, it is natural to consider their gauge covariant divergence.
After some algebra with commutators
\eqref{hatcomm} and Bianchi identities \eqref{bianchi-covariant},  we find 
\medskip
\begin{equation}
  \hat \nabla^\mu W_{\mu \nu}^{(i)} = \frac12 F^\mu{}_\nu B_\mu^{(i)}\, ,\qquad i=1,2,3,4,
  \,\, \textrm{fixed} \, .
\label{divW}
\end{equation} 
%
This result is due to the fact that  every integral $I_i$, is
separately invariant under dilatations. Then if
  the dilatation gauge field $\w_\mu$ is on-shell $B_\mu=0$, we have a conservation law
  \begin{align}
    \sum_{i=1}^4 a_i \hat \nabla^\mu W_{\mu \nu}^{(i)} = 0\, .
  \end{align}

\medskip\noindent
Therefore the energy-momentum tensor $W_{\mu \nu}$, eq.(\ref{eqs})
is Weyl gauge covariantly conserved
\medskip
\bea
\hat\nabla^\mu W_{\mu\nu}=0\, ,\qquad
W_{\mu\nu}
=\textrm{conserved}\, .
\eea

\medskip\noindent
This  conservation is consistent with our Weyl gauge covariant formalism and
our view of conformal geometry as a  physical  theory.
Note that the initial action is entirely a
geometric construct, including $\w_\mu$ as part of the Weyl connection that defines
this geometry, and with no matter fields or moduli fields  or other
non-geometric degrees of freedom.

Further, we consider the divergence of eqs. \eqref{B1} -- \eqref{B4}, to find
\medskip
\begin{align}
  \hat \nabla^\mu B_\mu^{(i)} = 2\, g^{\mu\nu}\,\,W_{\mu\nu}^{(i)}\, .
  \label{traceless}
\end{align}

\medskip\noindent
For $\w_\mu$ on-shell ($B_\mu=0$), we find the
  trace ($\tr\,W_{\mu \nu}\equiv g^{\alpha\beta} W_{\alpha\beta}$) of
energy momentum tensor 
\begin{equation}\label{traceless2}
  \tr\, W_{\mu \nu} = 0 \, .
\end{equation}

We would like a more intuitive picture of the conservation law for the
energy momentum tensor, from a  Riemannian geometry perspective
(in which $\w_\mu$ appears as a field ``added'' to this geometry,
not part of it). To this purpose we express
$\hat\nabla^\mu W_{\mu\nu}$ in terms of the
covariant derivative ($\mathring \nabla$) associated to the Levi-Civita
connection $\mathring \Gamma$. Since $\sqrt{g}\,  W_{\mu \nu} \delta g^{\mu \nu}$
is invariant it  follows that the space-time charge of $W_{\mu\nu}$
is $\tilde q_{W_{\mu\nu}} = -2$, therefore
\medskip
\begin{equation}
  \hat \nabla^\mu W_{\mu \nu}^{(i)} = \mathring \nabla^\mu W_{\mu \nu}^{(i)} - \w_\nu \, g^{\lambda\rho}
 \, W_{\lambda\rho}^{(i)}\, .
\label{divW-2}
\end{equation}

\medskip\noindent
Then, using the equation of motion for $\w_\mu$, ($B_\mu\!=\!0$, eq.(\ref{eqs})), and eq.~\eqref{traceless2}, the gauge covariant  law \eqref{divW} implies 
\begin{align}
  \mathring \nabla^\mu W_{\mu \nu} = 0  \, .
\end{align}

\medskip
Therefore, the energy-momentum tensor ($\propto W_{\mu \nu}$) is
also conserved with respect to  the covariant derivative
associated to the Levi-Civita connection.
In conclusion, $W_{\mu\nu}$ is also conserved from the Riemannian perspective.
This  is an interesting result that will be explained shortly
from a more general symmetry argument.

Consider now the Weyl gauge current associated to the dilatation symmetry
of action (\ref{action}). This can be  found  from the
equation of motion of $\w_\mu$.
In the basis of curvature operators of (\ref{action})
of covariant formulation, the Weyl-tensor-squared term is independent
of $\w_\mu$, while the Euler term does not affect it.
The equation of motion of $\w_\mu$  ($B_\mu=0$, eq.(\ref{eqs}))
can then be re-written in the form
 \begin{equation}
 4 \alpha_3 \hat \nabla_\nu F^{\nu}{}_\mu = j_\mu\, ,
 \label{def-current}
 \end{equation} 
with the notation
 \begin{equation}
j_\mu = 12 \alpha_1 \hat \nabla_\mu R \,.
\label{current}
 \end{equation}

 \medskip\noindent
 Taking the divergence of \eqref{def-current}, we see
 from \eqref{id-four} that,
 $j_\mu$ is gauge covariantly conserved
\begin{align}
 \hat \nabla^\mu j_\mu = 0 \, .
\end{align}

\medskip\noindent
Note that this is valid for $\w_\mu$ on-shell. Further, from eqs.(\ref{W1}) to (\ref{W4})
\medskip
\begin{align}
  \tr\, W_{\mu\nu}^{(1)} &= 6\, \hat \Box R\, ,
  & \tr\, W_{\mu\nu}^{(2)} &= 2\, \hat \Box R\, , &\tr\, W_{\mu\nu}^{(3)} &
  = 0\, , & \tr\,W_{\mu\nu}^{(4)} &= 2\, \hat \Box R\, . \label{trW}
\end{align}

\medskip\noindent
Thus $\tr\, W_{\mu\nu} = \sum_{i=1}^4 a_i \tr\, W_{\mu\nu}^{(i)} = (6a_1 + 2 a_2 + 2a_4) \hat \Box R
= 6 \alpha_1 \hat \Box R = \alpha_1 \tr\, W_{\mu\nu}^{(1)}$.
Therefore
\medskip
\begin{equation}
 2\, \tr\, W_{\mu\nu}\, = \hat \nabla^\alpha j_\alpha\, .
\end{equation}

\medskip\noindent
This result is  more general as it can be seen 
from a variation of the general action that is Weyl gauge invariant\footnote{See later, eqs.\eqref{dilat-1}, \eqref{dilatation}.}.
Further, since the space-time charge of $R$ is $\tilde q_R=-2$
one shows 
\medskip
\begin{equation}
\hat \Box R = \mathring \nabla_\mu \hat \nabla^\mu R\, .
\end{equation}
As a result
\begin{align}
\mathring \nabla^\mu j_\mu = 0 \, .
\end{align}

\medskip\noindent
Therefore, the conservation of the Weyl gauge current
is also  true with respect  to the Riemannian geometry/connection,
in agreement with \cite{Ghilen0,SMW}. This conserved current is a generalisation of a 
global scale symmetry current \cite{F1,F2,F3,F4,F5}.

Finally, let us consider the case of conformal gravity, usually
regarded as a gauge theory of conformal group \cite{Kaku}.  This action
is obtained by setting $\alpha_1=\alpha_3=0$ in action (\ref{action}). In this case
the Weyl current, hereafter denoted $j_\mu^{(c)}$,
is identically zero ($\alpha_1 = 0$) and  the associated tensor
$W_{\mu \nu}^{(c)}=2 [W_{\mu\nu}^{(2)}- 2 W_{\mu\nu}^{(3)}-1/3 \, W_{\mu\nu}^{(1)}]$
is conserved with a vanishing trace
\medskip
\begin{align}
  j_\mu^{(c)} = 0\, , && \hat \nabla^\mu W_{\mu \nu}^{(c)} =
  \mathring \nabla^\mu W_{\mu \nu}^{(c)} = 0\, , && \tr\, W_{\mu\nu}^{(c)} = 0\, .
\end{align}

\medskip\noindent
This  recovers the result that in conformal gravity the Weyl current
and thus the associated charge are trivial \cite{review}. This is
related to the fact that the gauge bosons of dilatations ($\w_\mu$)
and of special conformal transformations are not dynamical and part of the
physical spectrum. For this reason
conformal gravity is not  a true gauge theory (of the conformal group).

In conclusion, 
the tensor $W_{\mu\nu}$, that generalises the Einstein tensor to
Weyl quadratic gravity, and the dilatation current $j_\mu$ are  conserved.
These conservation laws are valid  with respect to the Weyl gauge covariant
derivative $\hat \nabla$ and also with respect to the Riemannian derivative
$\mathring \nabla$. The trace of energy momentum tensor
($=2 \tr \,W_{\mu \nu}$ in our notation) equals  the divergence of the current. In 
conformal gravity the current and its charge are trivial.

\subsection{Gauge covariant diffeomorphisms and conservation laws}
\label{sec-diffeo}

In General Relativity the conservation of the Einstein tensor and of the (matter)
energy-momentum tensor (on-shell) are a consequence of diffeomorphism (or general
coordinate transformation) invariance. The symmetry group of Weyl quadratic
gravity is larger, consisting of diffeomorphisms and dilatations.
We examine in the following how these are related to the conservation
laws (with respect to Weyl and Riemannian covariant derivatives),
that we derived earlier from the equations of motion.

Under a simultaneous variation of the metric $\delta g_{\mu\nu}$ and of the Weyl
gauge boson $\delta \w_\mu$, the resulting variation of the Weyl gravity action (\ref{action})
has the general form
\begin{equation}
  \delta S = \int d^d x\, \sqrt{g} \left[- W^{\mu \nu} \delta g_{\mu \nu}
    + B^\mu \delta \w_\mu \right]\, ,\label{general-variation} 
\end{equation}
where the minus sign arises from \eqref{def-WB} since
$W^{\mu\nu}\delta g_{\mu\nu}=-W_{\mu\nu} \delta g^{\mu\nu}$,
($ \delta g_{\rho\nu}=-g_{\rho\alpha} g_{\nu\beta}\delta g^{\alpha\beta}$).
Consider first the effect of general coordinate transformations generated by
a vector $\xi^\mu$: $x^\mu\ra x^{\prime \mu}=x^\mu+ \xi^\mu(x)$.
The induced variations on the fields are
\medskip
\begin{align}
  \delta_\xi g_{\mu \nu} & =- \xi^\rho \partial_\rho g_{\mu \nu}
  - g_{\rho \nu} \partial_\mu \xi^\rho - g_{\mu \rho} \partial_\nu \xi^\rho =
  - \mathring \nabla_\mu \xi_\nu - \mathring \nabla_\nu \xi_\mu \, , \label{diffeo-1}\\
  \delta_\xi \w_\mu & = - \xi^\nu \partial_\nu \w_\mu - (\partial_\mu \xi^\nu) \w_\nu=
  - \xi^\nu \mathring \nabla_\nu \w_\mu - (\mathring \nabla_\mu \xi^\nu) \w_\nu \, , \label{diffeo-2}
\end{align}

\medskip\noindent
where we have expressed the usual Lie derivatives in terms of the covariant
derivative $\mathring\nabla$ associated to the Levi-Civita connection.
The transformations above leave the action
invariant. After replacing them in \eqref{general-variation} and an
integration by parts we find
\begin{equation}
  \delta_\xi S =  \int d^d x \, \sqrt{g} \left[- 2 \mathring \nabla_\mu W^{\mu \nu}
    + F_\mu{}^\nu B^\mu + \w^\nu \mathring \nabla_\mu B^\mu \right] \xi_\nu = 0\, ,
\end{equation}
therefore 
\begin{equation}
  \mathring \nabla_\mu W^{\mu \nu} = \frac12 F_\mu{}^\nu B^\mu
  +\frac12 \w^\nu \mathring \nabla_\mu B^\mu  \, ,
\label{diffeo}
\end{equation}

\medskip\noindent
which recovers the Riemannian conservation law of $\mathring\nabla_\mu W^{\mu \nu}=0$
(with $\w_\mu$ onshell, $B_\mu=0$).

In addition to diffeomorphisms, the action of Weyl quadratic gravity is
also invariant under (gauged) dilatations which at infinitesimal level act as follows,
see eq.(\ref{gauge-transformation})
\begin{align}
\delta_D g_{\mu \nu} &= 2 \lambda_D g_{\mu \nu}\, ,\\
\delta_D \w_\mu &= - \partial_\mu \lambda_D = - \mathring \nabla_\mu \lambda_D
= - \hat \nabla_\mu \lambda_D \, .
\end{align}
We expressed the gauge transformation in the second equation  both
in terms of the Riemannian derivative $\mathring \nabla$ and the Weyl gauge covariant
derivative $\hat \nabla$ \footnote{This is related to the conservation of the dilatation current
both with Weyl and Riemannian covariance.}. Replacing the variations above
in \eqref{general-variation} yields, after integration by parts
\begin{equation}
\begin{split}
  \delta_D S &=  \int d^d x \, \sqrt{g} \left[-2 g_{\mu\nu} W^{\mu\nu}
    + \mathring \nabla_\mu B^\mu \right] \lambda_D = \int d^d x \, \sqrt{g}
  \left[-2 g_{\mu\nu} W^{\mu\nu} + \hat \nabla_\mu B^\mu \right] \lambda_D\, ,
\end{split} \label{dilat-1}
\end{equation}
which implies  the trace identity of $W_{\mu \nu}$
which can also be written in terms of the current
\begin{equation}
  2 g_{\mu\nu} W^{\mu\nu}
  = \hat \nabla_\mu B^\mu = \mathring \nabla_\mu B^\mu=  \hat \nabla_\mu j^\mu
  = \mathring \nabla_\mu j^\mu\, .
\label{dilatation}
\end{equation}

\medskip\noindent
One can check that combining the invariances under diffeomorphisms and
dilatations expressed in \eqref{diffeo} and \eqref{dilatation},
we obtain the already derived dilatation covariant conservation law \eqref{divW}.
However, as we show below, this can
be derived directly from invariance under {\it gauged} covariant diffeomorphisms, which
have the form
\begin{align}
  \delta g_{\mu \nu} & = - \hat \nabla_\mu \xi_\nu - \hat \nabla_\nu \xi_\mu =
  - \mathring \nabla_\mu \xi_\nu - \mathring \nabla_\nu \xi_\mu
  - 2 g_{\mu \nu} \w^\rho \xi_\rho\nonumber \\
  & =  -\xi^\rho \partial_\rho g_{\mu \nu} - g_{\rho \nu} \partial_\mu \xi^\rho
  - g_{\mu \rho} \partial_\nu \xi^\rho - 2 g_{\mu \nu} \w^\rho \xi_\rho \label{var-metric}\, ,\\[6pt]
  \delta \w_\mu & =  - \xi^\nu \partial_\nu \w_\mu - (\partial_\mu \xi^\nu) \w_\nu
  + \partial_\mu (\w^\rho \xi_\rho)\nonumber\\
  &=- \xi^\nu(\mathring \nabla_\nu \w_\mu) - (\mathring \nabla_\mu \xi^\nu) \w_\nu
  + \partial_\mu (\w^\rho \xi_\rho) = F_{\mu \nu} \xi^\nu \label{var-gauge}\, .
\end{align}

\medskip\noindent
From these equations we can see that a gauged covariant diffeomorphism\footnote{ Note
that \eqref{var-metric}
corresponds  to a gauged translation of the metric after imposing the constraint
$R_{\mu \nu}(P^a) = 0$ determining the spin connection as derived in the Appendix.}
is a combined action of a diffeomorphism generated by $\xi$ together with a dilatation
gauge transformation with parameter $\lambda_D =- \w^\rho \xi_\rho$.
These transformations are manifestly gauge covariant.  Imposing this symmetry at the
level of the action gives, after integration by parts
\medskip
\begin{equation}
  \delta S = \int d^d x \, \sqrt{g}
  \left[- 2 \hat \nabla_\mu W^{\mu \nu} + F_\mu{}^\nu B^\mu \right] \xi_\nu\, .
\end{equation}

\medskip\noindent
We thus obtained \eqref{divW} as a consequence of
invariance under \eqref{var-metric}-\eqref{var-gauge}. 
 
In conclusion, having both  Weyl covariant and Riemannian conservation laws for
$W_{\mu \nu}$ is a consequence of having diffeomorphism and dilatation invariance separately;
therefore, the choice of basis between gauged covariant diffeomorphisms
\eqref{var-metric}-\eqref{var-gauge} or (standard) diffeomorphisms
\eqref{diffeo-1}-\eqref{diffeo-2} (together with dilatations) yields the gauge covariant
or the Riemannian conservation law, respectively. Furthermore, the Weyl
gauge covariant and Riemannian
conservation laws of the dilatation current can be traced to the fact that the gauge
parameter $\lambda_D$ has a vanishing Weyl weight (by definition), therefore the action
of $\mathring \nabla$ and $\hat \nabla$ on it coincide. Finally, the symmetry arguments
given in this subsection are independent of the space-time dimension and therefore the
conservation laws are valid in arbitrary dimension.

\subsection{Riemannian interpretation}

 So far  we have seen that in conformal geometry  it is very useful
 to work in a manifestly Weyl gauge  covariant way.
 Nevertheless, our results can also be presented 
in the  Riemannian geometry picture. We  consider first the case of the symmetric phase
 for the Weyl quadratic Lagrangian and
 then the spontaneously broken phase. In the latter,  one recovers Einstein's general
 relativity coupled to a massive vector boson ($\w_\mu$) and a dynamically generated
 (positive) cosmological  constant \cite{Ghilen0,SMW}.

\bigskip\noindent
{\bf $\bullet$ Symmetric phase}

\medskip\noindent
One can regard the Weyl quadratic gravity action of eq.(\ref{action}),
discussed so far in the gauge covariant  picture, as a  Riemannian quadratic
gravity coupled to  the vector boson of dilatations through an energy-momentum tensor
($=2 T_{\mu \nu}(\w)$ in our notation below).
 Therefore, unlike in conformal geometry (where $\w_\mu$ was actually part of it),
 in the Riemannian picture we regard $\w_\mu$ as a ``matter'' field.
 The dilatation covariant tensor $W_{\mu \nu}$ is then split in a sum of two
 non-covariant parts
 \bea
W_{\mu \nu} = \mathring W_{\mu \nu} + T_{\mu \nu}(\w) \, ,
\label{split-0}
\eea
$\mathring W_{\mu \nu}$ depends only on the metric and
gives the quadratic gravity part. Hence the equation of motion
of the gravitational field is written as
\begin{equation}
\mathring W_{\mu \nu} = - T_{\mu \nu}(\w)\, .
\end{equation}
The left-hand-side is given by
\begin{align}
\mathring W_{\mu \nu} = \alpha_1 \mathring W_{\mu \nu}^{(1)} + \alpha_2 \mathring W_{\mu \nu}^{(c)} \, ,
\end{align}
with the Riemannian counterpart $\mathring W_{\mu \nu}^{(1)} = 2 g_{\mu \nu} \mathring \Box \mathring R
- 2\mathring \nabla_\nu \mathring \nabla_\mu \mathring R + 2 \mathring R_{\mu \nu} \mathring R
- \frac12 g_{\mu \nu} \mathring R^2$ and
\begin{equation}
\begin{split}
  \mathring W^{(c)}_{\mu \nu}  & = 
  -\frac13 g_{\mu \nu} \mathring \Box \mathring R + 2 \mathring \Box \mathring R_{\mu \nu}
  - \frac43 \mathring R \mathring R_{\mu \nu} + \frac13 g_{\mu \nu} \mathring R^2
  + \frac43 \mathring \nabla_\mu \mathring \nabla_\nu \mathring R   
   \\[6pt]
  &
 - 4 \mathring \nabla_\rho \mathring \nabla_{(\mu} \mathring R_{\nu )}{}^\rho
  + 4 \mathring R_{\mu \rho} \mathring R_\nu{}^\rho - g_{\mu \nu}
  \mathring R_{\rho \sigma} \mathring R^{\rho \sigma} \, .
 \label{Wc0}
 \end{split}
\end{equation}
The exact form of $T_{\mu \nu}(\w)$ is found after some algebra (with 
 \eqref{nabla} and \eqref{box}) and it has a rather complicated form,
showing the limitations of working in a non-covariant picture:
\medskip
\begin{equation}
\begin{split}
  T_{\mu \nu}(\w)&=6\alpha_1 \left[12 \w^2 \w_{\mu } \w_{\nu } - 3  g_{\mu \nu } \w^4
    - 2\w^2 \mathring R_{\mu \nu } - 2 \w_{\mu } \w_{\nu } \mathring R +
    \w^2 g_{\mu \nu } \mathring R  -  g_{\mu \nu } \w^{\rho}  \mathring \nabla_{\rho} \mathring R\right.\\
    & +  \w_{\mu } \mathring \nabla_{\nu } \mathring R +  \w_{\nu } \mathring \nabla_{\mu } \mathring R
    - 2 \mathring R_{\mu \nu }  \mathring\nabla_{\rho}\w^{\rho}
    + 12 \w_{\mu } \w_{\nu } \mathring \nabla_{\rho}\w^{\rho}
    + 12 g_{\mu \nu } \w^{\rho} \w^{\sigma}  \mathring \nabla_{\sigma}\w_{\rho} \\
    &- 12 \w^{\rho} \w_{\nu } \mathring \nabla_{\mu }\w_{\rho}
    - 12 \w^{\rho} \w_{\mu } \mathring \nabla_{\nu }\w_{\rho}
    + 3 g_{\mu \nu } \mathring \nabla_{\rho}\w^{\rho} \mathring \nabla_{\sigma}\w^{\sigma}
    - 4 g_{\mu \nu } \mathring \nabla_{\sigma}\w_{\rho} \mathring \nabla^{\sigma}\w^{\rho}\\
    & + 4 \mathring \nabla_{\mu }\w^{\rho} \mathring \nabla_{\nu }\w_{\rho}
    - 4 g_{\mu \nu } \w^{\rho}  \mathring\Box \w_{\rho}
    - 6 \w_{\nu } \mathring \nabla_{\mu } \mathring \nabla_{\rho}\w^{\rho}
    - 6 \w_{\mu } \mathring \nabla_{\nu }\mathring \nabla_{\rho}\w^{\rho} \\
    &\left. + 6 g_{\mu \nu } \w^{\rho}  \mathring \nabla_{\rho} \mathring \nabla_{\sigma}\w^{\sigma}
    + 4 \w^{\rho} \mathring \nabla_{\nu } \mathring \nabla_{\mu }\w_{\rho}
    - 2 g_{\mu \nu } \mathring \Box \mathring \nabla_{\rho}\w^{\rho}
    + 2 \mathring \nabla_{\nu }\mathring \nabla_{\mu } \mathring \nabla_{\rho}\w^{\rho} \right]\\ 
  &+  \alpha_3 \Big[2 F_{\mu \rho} F_\nu{}^\rho
    - \frac12 g_{\mu \nu} F_{\rho \sigma} F^{\rho \sigma} \Big]\, .
\end{split}
\label{em-w}
\end{equation}

\medskip\noindent
To derive this equation one had to express $R^\rho{}_{\sigma \mu \nu}$ and its
contractions in terms of its Riemannian counterparts by
using eqs.(\ref{riem-1}), (\ref{riem-2}), (\ref{riem-3}).
Although \eqref{em-w} is rather complicated, from
formulae \eqref{divW}, \eqref{traceless} and \eqref{divW-2} 
derived in the covariant picture, one finds that we have
\medskip
\begin{align}
  \mathring \nabla^\mu \mathring W_{\mu \nu} =0\, , && \mathring \nabla^\mu T_{\mu \nu}(\w)
  = \frac12 \left(F^\mu{}_\nu B_\mu + \w_\nu \hat \nabla^\mu B_\mu \right) \, .
\end{align}

\medskip\noindent
Therefore, similar to  other cases  in general
relativity with matter present, the energy-momentum tensor is
conserved when the equation of motion of ``matter'' (in our case $\w_\mu$)
is satisfied ($B_\mu=0$). Then 
\begin{align}
  \mathring \nabla^\mu T_{\mu \nu}(\w) = 0 \, .
\end{align}
Consider also the equation $B_\mu = 0$ in the Riemannian interpretation.
It is useful to note  the following identity 
\medskip
\begin{equation}
\hat \nabla_\nu F^\nu{}_\mu = \mathring \nabla_\nu F^\nu{}_\mu\, ,
\end{equation} 
and therefore we can write 
\begin{equation}
4 \alpha_3 \mathring \nabla_\nu F^\nu{}_\mu = j_\mu\, ,
\end{equation}

\medskip\noindent
where the  current $j_\mu$ is  expressed with the help of \eqref{riem-3} as
\medskip
\begin{equation}
  j_\mu \sim \hat \nabla_\mu R = \mathring \nabla_\mu \mathring R
  - 2 \w_\mu \mathring R - 6 \mathring \nabla_\mu \mathring \nabla_\nu \w^\nu
  - 12 \w_\nu \mathring \nabla_\mu \w^\nu + 12 \w_\mu \mathring \nabla_\nu \w^\nu + 12 \w_\nu \w^2 \, .
\end{equation}

\medskip\noindent
This reveals a non-Abelian structure of self-interactions for the dilatation gauge boson.
This is not surprising from the gauge theory point of view since the Weyl group is
actually a semi-direct product of the dilatations with the Poincar\'e group. 

\bigskip
\bigskip\noindent
 {$\bullet$ \bf Spontaneously broken phase and conservation laws}

 \medskip\noindent
 Consider now the spontaneously broken phase of Weyl quadratic gravity (\ref{action}),
 for details see  \cite{Ghilen0,SMW} (\cite{review} for a review) and
 present the conservation laws
 from a Riemannian perspective. We first show how to ``linearise'' a relevant term in the action
(which triggers the breaking), then analyse the general action and conservation laws.

 First,  one  linearises the $R^2$ term in the Lagrangian (\ref{action}) with
the help of a scalar field $\phi$. The corresponding term $R^2$
in the action will be written as
\medskip
\begin{equation}
\tilde I_1 = -\int d^4 x \sqrt{g} \, \left (2 \phi^2 R + \phi^4 \right)\, ,
\label{GR}
\end{equation}
which is   equivalent to $I_1$ of (\ref{eq}); this is seen by using the
equation of motion of $\phi$
\bea\label{eom-phi}
\phi^2=-R \, ,
\eea
where we assumed $\phi\not =0$. When this solution
is inserted back into $\tilde I_1$, one recovers $I_1$, hence their equivalence.
The Weyl weight of $\phi$ (as seen from the solution) is
$\tilde q_\phi = q_\phi = -1$, so the action above remains invariant under
gauged dilatations \eqref{gauge-transformation} when $\phi$ transforms as
\begin{equation}
\phi \mapsto e^{- \lambda_D} \phi \, .
\end{equation}
Varying $\tilde I_1$ with respect to $g_{\mu\nu}$  and the Weyl
gauge boson $\w_\mu$ one obtains $\tilde W_{\mu\nu}^{(1)}$ and $\tilde B_\mu^{(1)}$
(which are counterparts to those in \eqref{def-WB}, \eqref{eqs2})
and contribute to the equations of motion
\begin{align}
  \tilde W_{\mu \nu}^{(1)} &= -2 g_{\mu \nu} \hat \Box \phi^2
  + 2 \hat \nabla_\nu \hat \nabla_\mu\phi^2 - 2 R_{\mu \nu} \phi^2
  + g_{\mu \nu} \phi^2 R + \frac12 g_{\mu \nu} \phi^4\, , \label{W-tilde}\\
\tilde B_\mu^{(1)} & = -12 \hat \nabla_\mu \phi^2\, . \label{B-tilde}
\end{align}

\medskip\noindent
These are equal to $W_{\mu \nu}^{(1)}$ and $B_\mu^{(1)}$, respectively,  upon use
of the equation of motion  $\phi^2 = -R$.
The analogue of identities \eqref{divW} and \eqref{diffeo}
for the tensor $\tilde W_{\mu \nu}^{(1)}$ are found to be \footnote{For the trace one obtains $
  \tr\,\tilde W^{(1)} = \frac12 \hat \nabla_\mu \tilde B^{(1)\mu}
  + 2 \phi^2 (R+ \phi^2) $.}
\begin{align}
  \hat \nabla^\mu \tilde W_{\mu \nu}^{(1)} &= \frac12 F_{\mu \nu} \tilde B^{(1)\mu}
  + 2 \phi(R+ \phi^2)\, \hat \nabla_\nu \phi\, , \label{law-1}\\
   \mathring \nabla^\mu \tilde W_{\mu \nu}^{(1)} &= \frac12 F_{\mu \nu} \tilde B^{(1)\mu}
  + \frac12 \w_\nu \mathring \nabla_\mu \tilde B^{(1)\mu}
  + 2 \phi (R+ \phi^2) \mathring \nabla_\nu \phi \, . \label{law-2}
\end{align}

The theory of (\ref{action}) in  its  ``linearised'' equivalent  version   
(with $\tilde I_1$ replacing $I_1$) has now  the equations of motion
given by the vanishing of the 
tensors $\tilde W_{\mu \nu}$ and $\tilde B_\mu$, where
\medskip
\begin{align}
  \tilde W_{\mu \nu} & = \alpha_1 \tilde W_{\mu \nu}^{(1)} + \alpha_2 W_{\mu \nu}^{(c)}
  + \alpha_3 W_{\mu \nu}^{(3)} \, , & \tilde B_\mu =
  \alpha_1 \tilde B_\mu^{(1)} + \alpha_3 B_\mu^{(3)} \, ,
\label{linear-weyl}
\end{align}

\medskip\noindent
together with the equation of motion of $\phi$ in \eqref{eom-phi}.
The  tensor $W_{\mu \nu}^{(c)}$ is exactly equal to that
given in eq.\,\eqref{Wc0}\footnote{See also \eqref{Wc}.}.
With the help of \eqref{law-1}-\eqref{law-2} one finds that $\tilde W_{\mu \nu}$ is conserved 
\medskip
\begin{equation}
\hat \nabla^\mu \tilde W_{\mu \nu} = \mathring \nabla^\mu \tilde W_{\mu \nu} = 0\, ,
\end{equation}

\medskip\noindent
when {\it both} $\w_\mu$ and $\phi$ are onshell.
The conserved current is now written as $j_\mu = -12 \alpha_1 \hat \nabla_\mu \phi^2$.

We can now address the Riemannian interpretation of 
Weyl quadratic gravity in this  ``linearised'' version, first in its symmetric and then
in its broken phase.
The breaking of Weyl gauge symmetry
is obvious by setting $\phi=\langle\phi\rangle$ (non-zero vacuum expectation value)
and then the Einstein-Hilbert term is generated by the first term in $\tilde I_1$
\cite{Ghilen0,SMW}\footnote{The field $\log\phi$  plays the role of the ``dilaton'';
more exactly $\log\phi$ is the
would-be-Goldstone of Weyl gauge symmetry which is eaten by $\w_\mu$ that becomes
massive in a Stueckelberg mechanism \cite{Ghilen0,SMW}.}.

We are first  looking
for a separation similar to \eqref{split-0}. An obvious choice is to treat both
$\w_\mu$ and $\phi$ as ``matter'' fields (in a  Riemannian interpretation)
and define an energy-momentum tensor
$T(\w,\phi)$ containing all the terms depending on these fields.
We thus  write
\begin{equation}
\tilde W_{\mu \nu} = \alpha_2 W_{\mu \nu}^{(c)}(g) + T_{\mu \nu}(\w,\phi)\, ,
\end{equation}

\medskip\noindent
where we emphasised that $W_{\mu \nu}^{(c)}$ depends only on the metric.
From \eqref{linear-weyl} it follows that the expression of the energy-momentum tensor is
\medskip
\begin{equation}
T_{\mu \nu}(\w,\phi) = \alpha_1 \tilde W_{\mu \nu}^{(1)} + \alpha_3 W_{\mu \nu}^{(3)}\, .
\end{equation}   

\medskip\noindent
The definition above has the virtue of having a dilatation covariant energy-momentum
tensor which satisfies both the Weyl gauge  covariant and the  Riemannian conservation laws when
{\it both} $\w_\mu$ and $\phi$ are on-shell
\begin{align}
  \tilde B_\mu &= 0 \quad \textrm{and}
  \quad \phi^2= - R  && \Ra & \hat \nabla^\mu  T_{\mu \nu}(\w,\phi)
  &= \mathring \nabla^\mu T_{\mu \nu}(\w,\phi) = 0 \, .
\end{align} 

\medskip
A different definition of the energy-momentum tensor is more useful when
considering the broken phase of the theory. In this case we adopt the following
separation
\medskip
\begin{equation}\label{s2}
\tilde W_{\mu \nu} = \mathring W_{\mu \nu}(g,\phi) + \tilde T_{\mu \nu}(\w,\phi)\, ,
\end{equation}
where some terms involving the scalar field $\phi$ are kept in the gravity part;
we  introduced
\medskip
\begin{align}
\mathring W_{\mu \nu}(g,\phi) & = -2 \alpha_1 \phi^2 \mathring G_{\mu \nu} + \alpha_2 W_{\mu \nu}^{(c)} \, ,
\end{align}
with
\bea
\mathring G_{\mu \nu} = \mathring R_{\mu \nu} - \frac12 g_{\mu \nu} \mathring R\, ,
\eea
for the usual (Riemannian) Einstein tensor, thus justifying our definition.
The matter energy-momentum tensor is
\begin{equation}
\begin{split}
  \tilde T_{\mu \nu} (\w,\phi) &=\alpha_1 \Big[-2 g_{\mu \nu} \mathring \Box \phi^2
    + 2 \mathring \nabla_\nu \mathring \nabla_\mu \phi^2 + 6(g_{\mu \nu}\w^\rho \mathring \nabla_\rho \phi^2-\w_\mu \mathring \nabla_\nu \phi^2 - \w_\nu \mathring \nabla_\mu \phi^2)
 \\
    & 
    -6g_{\mu \nu}\w^2 \phi^2 + 12 \w_\mu \w_\nu \phi^2 + \frac12 g_{\mu \nu} \phi^4    \Big]
  +  \alpha_3 \Big[2 F_{\mu \rho} F_\nu{}^\rho
    - \frac12 g_{\mu \nu} F_{\rho \sigma} F^{\rho \sigma} \Big] \, .
\end{split}
\end{equation}
Note that neither $\mathring W_{\mu \nu}(g,\phi)$ nor $T_{\mu \nu}(\w,\phi)$ of \eqref{s2}
are dilatation covariant - this is not required from a Riemannian view. Their
conservation laws are valid only in the broken phase, when the scalar field is
replaced by its vacuum expectation value; indeed, with
\medskip
\begin{align}
  \tilde B_\mu = 0, \,\,\,
  R+\phi^2=0\,\,\textrm{and}\,\,
  \phi = \langle \phi \rangle
  \,\,\,\Ra \,\,\, \mathring \nabla^\mu \mathring W_{\mu \nu}(g,\phi)
  =0 \quad \textrm{and} \quad \mathring \nabla^\mu \tilde T_{\mu \nu}(\w,\phi)  = 0 \, .
\end{align}

\medskip
As already mentioned, in  the broken phase, with $\langle\phi\rangle\not=0$, 
as seen from  \eqref{GR}  one generates an Einstein-Hilbert term in the
action together with a (positive) cosmological constant. Upon expanding $R$ in
terms of its Riemannian part eq.\eqref{riem-3}, one can see that a mass term
is generated for the dilatation gauge field. Hence, in the
spontaneous broken phase one recovers general relativity (with dynamically
generated Planck mass and cosmological constant)  coupled to a massive vector
field \cite{Ghilen0,SMW} and a Weyl tensor-squared term.

\section{Some examples}\label{5}

To illustrate the previous results in conformal geometry, we present some particular
cases of the general action (\ref{action}) and equations of motion
(\ref{eom-1}), (\ref{eom-2}), corresponding to: 
Gauss-Bonnet gravity, $R^2$-Weyl gravity and conformal gravity.

\bigskip\noindent
{\bf $\bullet$ Gauss-Bonnet gravity in  conformal geometry}

\medskip\noindent
We first consider the non-dynamical (in four dimensions)
Gauss-Bonnet gravity described by the action below, in the
  Weyl gauge covariant formulation of conformal geometry:
\begin{equation}
\begin{split}
S_G &= \int d^4 x\, \sqrt{g}\, \left[R^2 - 4 R_{\mu \nu} R^{\nu \mu} + R_{\mu \nu \rho \sigma} R^{\rho \sigma \mu \nu} \right]\\
& = \int d^4 x \, \sqrt{g} \left[R^2 - 4 R_{\mu \nu} R^{\mu \nu} + 4 F_{\mu \nu} F^{\mu \nu} + R_{\mu \nu \rho \sigma} R^{\mu \nu \rho \sigma} \right] \, .
\end{split} 
\end{equation}
As seen from this action, the equations of motion of Gauss-Bonnet gravity are a particular linear
combination of $W_{\mu \nu}^{(i)}$ for the metric and the same linear combination
of $B_\mu^{(i)}$ for the Weyl gauge field. Then let us denote
\begin{align}
  W_{\mu \nu}^{(G)} &: = W_{\mu \nu}^{(1)} - 4 W_{\mu \nu}^{(2)} + 4 W_{\mu \nu}^{(3)}
  + W_{\mu \nu}^{(4)}\, ,\\
B_\mu^{(G)} & : = B_{\mu}^{(1)} - 4 B_{\mu}^{(2)} + 4 B_{\mu}^{(3)} + B_{\mu}^{(4)}\, ,
\end{align}

\medskip\noindent
where the combination above is different from  the usual Riemannian one
due to the presence of torsion in the curvature tensor. It is easy to check
that $B_\mu^{(G)} \equiv 0$ from eqs.\,\eqref{B1}-\eqref{B4} and thus the
corresponding equation of motion is trivial, as expected.
Regarding $W_{\mu \nu}^{(G)}$ we simplify it using commutators of covariant
  derivatives and Bianchi identities. For example, the first term in $W_{\mu\nu}^{(4)}$,
  eq.\eqref{W4} is computed using \eqref{Euler-id} while remaining terms
  are expressed in terms of $C_{\mu\nu\rho\sigma}^2$, eq.(\ref{weyl-squared}). One finds
\begin{equation}
  W^{(G)}_{\mu \nu} = 2 C_{\rho \sigma \alpha \mu} C^{\rho \sigma \alpha}{}_\nu
  - \frac12 g_{\mu \nu} C_{\rho \sigma \alpha \beta} C^{\rho \sigma \alpha \beta} \equiv 0\, ,
\label{GB}
\end{equation}
where the last equality corresponds to a well-known identity for
the Weyl tensor valid  in four dimensions only (see \cite{Lovelock}
for a proof). Notice that it also applies to our case since we have
$C_{\mu \nu \rho \sigma} = \mathring C_{\mu \nu \rho \sigma}$. The result above provides
a non-trivial check of our formulae.\\

\bigskip\noindent
{\bf $\bullet$ $R^2 + F^2$-Weyl gravity}

\bigskip\noindent
A simpler case than the general action (\ref{action}) is when
the Weyl-tensor-squared term is not present in the action.
Consider then the following  action in Weyl conformal geometry
\begin{equation}\label{SW}
S_w = \int d^4 x \, \sqrt{g} \left[\alpha_1 R^2 + \alpha_3 F_{\mu \nu} F^{\mu \nu} \right] \, .
\end{equation}

The equations of motion in this case are found by setting
$\alpha_2 = \alpha_4 = 0$ (equivalently $a_2 = a_4 = 0$) in the general
action. We then have
\medskip
\begin{align}\label{RF}
  W_{\mu \nu}^{(w)} &= \alpha_1 W_{\mu \nu}^{(1)} + \alpha_3 W_{\mu \nu}^{(3)}\, , & B_\mu^{(w)} &
  = \alpha_1 B_\mu^{(1)} + \alpha_3 B_{\mu}^{(3)}\, .
\end{align}

\medskip\noindent
From (\ref{eom-1}), (\ref{eom-2})  one finds the dynamical equations
\medskip
\begin{align}
  \alpha_1 \Big(2 g_{\mu \nu} \hat \Box R - 2\hat \nabla_\nu \hat \nabla_\mu R
  + 2 R_{\mu \nu} R - \frac12 g_{\mu \nu} R^2 \Big) + \alpha_3 \Big(2 F_{\mu \rho} F_\nu{}^\rho
  - \frac12 g_{\mu \nu} F_{\rho \sigma} F^{\rho \sigma} \Big) &= 0\, ,\\
 12 \alpha_1 \hat \nabla_\mu R - 4 \alpha_3 \hat \nabla_\nu F^\nu{}_\mu &= 0\, .
\end{align}

\medskip\noindent
The last equation indicates that we have a conserved current of the
form 
\medskip
\begin{equation}
j_\mu^{(w)} = 12 \alpha_1 \hat \nabla_\mu R \, .
 \end{equation}

\medskip\noindent
This current  is the same as in the general case in eq.\,\eqref{current}
since the Weyl-tensor-squared term (that we did not include  in the action here)
does not contribute to it. \\

\bigskip\noindent
  {\bf $\bullet$ $R^2$-Weyl gravity}

  \bigskip\noindent
  Consider now the previous case, eq.(\ref{SW}), when the Weyl gauge boson is not dynamical,
  which means $F_{\mu\nu}=0$ in $S_w$.
  The action simplifies to
\begin{equation}
S_R = \int d^4 x \, \sqrt{g}\,  \alpha_1 R^2 \, .
\end{equation} 
Setting $\alpha_3=0$ in eq.(\ref{RF}) then
\begin{align}
W_{\mu \nu}^{(R)} &= \alpha_1 W_{\mu \nu}^{(1)}\, , & B_\mu^{(R)} = \alpha_1 B_\mu^{(1)} \, .
\end{align}
The equations of motion are then
\begin{align}
  2 g_{\mu \nu} \hat \Box R - 2\hat \nabla_\nu \hat \nabla_\mu R
  + 2 R_{\mu \nu} R - \frac12 g_{\mu \nu} R^2  &= 0\, ,\\
  \hat \nabla_\mu R & = 0 \, .
\end{align}
Notice that the second equation is equivalent
to having the corresponding current $j_\mu^{(R)} = 0$.
This equation can, equivalently, be written as
\begin{equation}\label{JR}
\w_\mu = \frac12 \partial_\mu \log |R| \, ,
\end{equation} 
for $R \neq 0$, showing that $\w_\mu$ is ``pure gauge'', as expected from our $F_{\mu\nu}=0$ condition.
Moreover, in the ``linearised'' version with an action
$\tilde S_R = \alpha_1 \tilde I_1$ with $\tilde I_1$ given in
eq.\,\eqref{GR} the gauge field of dilatations satisfies
$\w_\mu =  (1/2)\,\partial_\mu \log \phi^2$, consistent with (\ref{JR})
for $\phi^2=-R$.

\bigskip\noindent
{\bf $\bullet$ Conformal gravity}\\

\noindent
The final example  that we consider in conformal geometry  is conformal gravity, defined
by an action containing only the Weyl tensor 
\begin{equation}
S_c = \int d^4 x \sqrt{g}\,\alpha_2\,  C_{\mu \nu \rho \sigma} C^{\mu \nu \rho \sigma} = \int d^4 x \, \sqrt{g}\, \alpha_2\,\left[\frac13 R^2 - 2 R_{\mu \nu} R^{\nu \mu} + R_{\mu \nu \rho \sigma} R^{ \rho \sigma \mu \nu} \right]\, .
\end{equation}

\medskip\noindent
Using that the Euler term does not contribute to the equations of motion, we can  write
\medskip
\begin{equation}
  S_c = \int d^4 x\,  \sqrt{g}\, \alpha_2 \Big[
    - \frac23 R^2 + 4 F_{\mu \nu} F^{\mu \nu } + 2 R_{\mu \nu} R^{\nu \mu}\Big] =
  \int d^4 x\, \sqrt{g}\, \alpha_2\Big[
    - \frac23 \mathring R^2
    + 2 \mathring R_{\mu \nu} \mathring R^{ \mu \nu}\Big] \, . \label{action-conformal}
\end{equation}

\medskip\noindent
In the last step we used eqs.\eqref{riem-2}, \eqref{riem-3}.
Equivalently, we could have used
that $C_{\mu \nu \rho \sigma} = \mathring C_{\mu \nu \rho \sigma}$, eq.(\ref{Cw}).
From this action, in the basis $W_{\mu \nu}^{(i)}$, $B_\mu^{(i)}$,
the equations of motion of conformal gravity are expressed  as the vanishing
of $W_{\mu\nu}^{(c)}$ and $B_\mu^{(c)}$, where
\begin{align}
  W_{\mu \nu}^{(c)} = 2 
  \Big[W_{\mu \nu}^{(2)}-2 W_{\mu \nu}^{(3)}
    - \frac13 W_{\mu \nu}^{(1)} \Big] \, ,
  && B_\mu^{(c)} = 2 
  \Big[B_{\mu}^{(2)}-2 B_{\mu}^{(3)} - \frac13 B_{\mu}^{(1)} \Big]\, .
\end{align}
The equation of motion of the metric $W_{\mu\nu}^{(c)}=0$
can be written in a manifestly dilatation covariant form
\begin{align}
  W_{\mu \nu}^{(c)} & = 
   -\frac13 g_{\mu \nu} \hat \Box R + 2 \hat \Box R_{(\mu \nu)}
  - 4 \hat \nabla_\rho \hat \nabla_{(\nu} g_{\mu) \sigma} R^{(\rho \sigma)}
  + \frac43 \hat \nabla_\nu \hat \nabla_\mu R
  + 4 R_{(\mu \rho)} R_{(\nu \sigma)} g^{\rho \sigma} 
  \nonumber\\
  &
  -g_{\mu \nu}R_{(\rho \sigma)} R^{(\rho \sigma)} - \frac43 R_{\mu \nu} R
  + \frac13 g_{\mu \nu} R^2 - 4 F_{\mu \rho} F_\nu{}^\rho
  + g_{\mu \nu} F_{\rho \sigma} F^{\rho \sigma} \, =0.
\label{conformal-covariant}
\end{align}
The equation of motion of the Weyl gauge boson is trivially respected with eqs.(\ref{B1}) to
(\ref{B3}), illustrating  that $\w_\mu$ is not dynamical in conformal gravity
\begin{equation}
B_\mu^{(c)} = 0 \, .
\end{equation} 
This is  obvious from action \eqref{action-conformal} which is independent of $\w_\mu$.

One can also check that in \eqref{conformal-covariant} the dependence of
$\w_\mu$ cancels out, therefore
\begin{equation}
\begin{split}
  W_{\mu \nu}^{(c)} =\mathring W^{(c)}_{\mu \nu}  &=
  -\frac13 g_{\mu \nu} \mathring \Box \mathring R + 2 \mathring \Box \mathring R_{\mu \nu}
  - \frac43 \mathring R \mathring R_{\mu \nu} + \frac13 g_{\mu \nu} \mathring R^2
  + \frac43 \mathring \nabla_\mu \mathring \nabla_\nu \mathring R \\
  & - 4 \mathring \nabla_\rho \mathring \nabla_{(\mu} \mathring R_{\nu )}{}^\rho
  + 4 \mathring R_{\mu \rho} \mathring R_\nu{}^\rho - g_{\mu \nu}
  \mathring R_{\rho \sigma} \mathring R^{\rho \sigma} \, .
\end{split}
\label{Wc}
  \end{equation}
Hence,  the equation of motion $W_{\mu\nu}^{(c)}=0$
recovers the well known equations of conformal gravity depending only
on Riemannian quantities but in non-covariant form \cite{mannheim}.
Finally, the dilatation current is identically zero in this case $j_\mu^{(c)}\equiv 0$.

\section{Conservation laws and equations of motion in $d$ dimensions}

In this section we generalise the equations of motions
and conservation laws that we derived so far,
to   action (\ref{action-d}) of conformal geometry in  $d$ dimensions,
{in the Weyl gauge covariant formulation}.
The formalism below is relevant for the quantum corrected action, when calculations
are performed in $d=4-2\epsilon$ dimensions\footnote{Note that $\epsilon$ is {\it not}
assumed to be {\it small} hence our results are also valid for models with extra dimensions.}. 
We thus derive these equations for general Weyl quadratic gravity in {\it arbitrary}
$d$ dimensions, while maintaining manifest Weyl gauge  {invariance}
of the {regularised} action in $d$ dimensions.
This is indeed possible and is important in order for the action to remain
Weyl-anomaly-free \cite{DG1,review}. 

\subsection{Equations of motion}

For convenience let us write below  action (\ref{action-d})
\begin{equation}
  S = \int d^d x \sqrt{g}\, \left[\alpha_1 R^2
    + \alpha_2 C_{\mu \nu \rho \sigma} C^{\mu \nu \rho \sigma}
    + \alpha_3 F_{\mu \nu} F^{\mu \nu} + \alpha _4 G \right]R^{(d-4)/2} \, ,
\label{action-d2}
\end{equation}
%
$S$ is indeed Weyl gauge invariant in $d$ dimensions.
The usual ultraviolet  regulator scale ({$\mu^{-\epsilon}$})
is replaced here by the Weyl scalar curvature $R^{-\epsilon}$ which transforms
covariantly  (\ref{cov1}) and, therefore, the action remains Weyl gauge invariant
and with dimensionless couplings.

Similar to the case in four dimensions, we define the integrals $I_i(\epsilon)$ for
$d=4-2\epsilon$
\begin{align}
  I_1(\epsilon) &= \int d^d x \sqrt{g}\, R^{2- \epsilon}\, , & I_2(\epsilon) &
  = \int d^d x \sqrt{g}\,R_{\mu \nu} R^{\mu \nu} R^{-\epsilon}\, ,\\
  I_3(\epsilon) & = \int d^d x \sqrt{g}\,F_{\mu \nu } F^{\mu \nu} R^{-\epsilon}\, ,
  & I_4(\epsilon) & = \int d^d x \sqrt{g}\,R_{\mu \nu \rho \sigma }
  R^{\mu \nu \rho \sigma} R^{-\epsilon} \, ,
\end{align}

\medskip\noindent
such that $I_i(0) = I_i$ of (\ref{eq}).
Their variations are
\medskip
\begin{align}
  \delta I_i(\epsilon) &= \int d^d x \sqrt{g} \,
  W_{\mu \nu}^{(i)}(\epsilon)\, \delta g^{\mu \nu}\, ,
  & \delta_\w I_i(\epsilon)
  & = \int d^d x \sqrt{g} \, B^{(i)\, \mu}(\epsilon)\, \delta \w_\mu \, ,
\end{align}

\medskip\noindent
with relations below that  generalise (\ref{a-alpha}) to $d$ dimensions
\begin{align}
  a_1 &= \alpha_1 + \frac{2 \alpha_2}{(d-1)(d-2)} + \alpha_4\, , & a_2 &
  = - \frac{4 \alpha_2}{d-2} - 4 \alpha_4\, , \nonumber\\
  a_3 & = \alpha_3+2(d-2)(d-3)\alpha_4\, , & a_4 &=
  \alpha_2 + \alpha_4\, . \label{a-alpha-d}
\end{align}
These are used when going from the basis of operators present
in action \eqref{action-d2} to that defined by the integrals above, via
(\ref{eq32}). Eq.(\ref{eqs2}) is modified into
 \bea\label{eqs2-epsilon}
 W_{\mu\nu}(\epsilon)\equiv \frac{1}{\sqrt{g}}
 \frac{\delta S}{\delta g^{\mu\nu}}=\sum_{i=1}^4 a_i \,W_{\mu\nu}^{(i)}(\epsilon)\, ,\qquad
 \textrm{and}\qquad
 B_\mu(\epsilon)\equiv \frac{g_{\mu\nu}}{\sqrt{g}} 
\frac{\delta S}{\delta \w_\nu}=\sum_{i=1}^4 a_i \,B_\mu^{(i)}(\epsilon)\, .
 \eea
 The equation of motion  for the metric is then
 $W_{\mu\nu}(\epsilon)=0$ where
 \medskip
 \begin{align}
  W_{\mu \nu}^{(1)}(\epsilon) & = 2  g_{\mu \nu} \hat \Box R^{1- \epsilon}
  - 2 \hat \nabla_{\nu} \hat \nabla_\mu R^{1- \epsilon} + 2 R_{\mu \nu} R^{1- \epsilon}
  - \frac12 g_{\mu \nu} R^{2- \epsilon} \nonumber\\
  & - \epsilon \left[- \hat \nabla_{\nu} \hat \nabla_{\mu} + R_{\mu \nu}
    + g_{\mu \nu} \hat \Box  \right] (R^{-\epsilon-1} R^2)\, ,\\[5pt]
  W_{\mu \nu}^{(2)}(\epsilon) & =
  g_{\mu \nu} \hat \nabla_\rho \hat \nabla_\sigma (R^{-\epsilon} R^{\rho \sigma})
  + \hat \Box (R^{-\epsilon} R_{(\mu \nu)}) -
  \hat \nabla_\sigma \hat \nabla_{(\nu} g_{\mu) \rho} (R^{-\epsilon} R^{\rho \sigma})\nonumber\\
  &  - \hat \nabla_\rho \hat \nabla_{(\nu} g_{\mu) \sigma} (R^{-\epsilon} R^{\rho \sigma})
  + R^{-\epsilon}(R_{\mu \rho} R_{\nu \sigma} + R_{\rho \mu} R_{\sigma \nu}) g^{\sigma \rho}
  - \frac12 g_{\mu \nu} R^{-\epsilon} R_{\rho \sigma} R^{\rho \sigma} \nonumber\\
  & - \epsilon \left[- \hat \nabla_{\nu} \hat \nabla_{\mu} + R_{\mu \nu}
    + g_{\mu \nu} \hat \Box  \right] (R^{-\epsilon-1} R_{\rho \sigma} R^{\rho \sigma}) \, ,\\
  W_{\mu \nu}^{(3)}(\epsilon) & = R^{-\epsilon} \left[2 F_{\mu \rho} F_\nu{}^\rho
    - \frac12 g_{\mu \nu} F_{\rho \sigma} F^{\rho \sigma} \right]
  - \epsilon \left[R_{\mu\nu}- \hat \nabla_{\nu} \hat \nabla_{\mu} 
    + g_{\mu \nu} \hat \Box  \right] (R^{-\epsilon-1} F_{\rho \sigma} F^{\rho \sigma}) ,\\
  W_{\mu \nu}^{(4)}(\epsilon) & =
  - 4 \hat \nabla_\sigma \hat \nabla_\rho (R^{-\epsilon} R^\sigma{}_{(\mu \nu)}{}^ \rho)
  + 2 R^{-\epsilon} R_{\rho \sigma \alpha \mu} R^{\rho \sigma \alpha}{}_\nu
  - \frac12 g_{\mu \nu} R^{-\epsilon} R_{\rho \sigma \alpha \beta} R^{\rho \sigma \alpha \beta} \nonumber\\
  & - \epsilon \left[- \hat \nabla_{\nu} \hat \nabla_{\mu} + R_{\mu \nu}
    + g_{\mu \nu} \hat \Box  \right] (R^{-\epsilon-1} R_{\rho \sigma \alpha \beta} R^{\rho \sigma \alpha \beta}) \, .
\end{align}

\medskip
Similar to  the four dimensional case, some symmetrisations over pairs of indices are
not necessary because of the identity $- \hat \nabla_{[\nu} \hat \nabla_{\mu]} R^{1-\epsilon}
+ R_{[\mu \nu]} R^{1- \epsilon} = 0$
and similar ones with $R^{1- \epsilon}$ replaced by
$R^{-\epsilon-1} R_{\rho \sigma} R^{\rho \sigma}$, $R^{-\epsilon-1} F_{\rho \sigma} F^{\rho \sigma}$
and $R^{-\epsilon-1} R_{\rho \sigma \alpha \beta} R^{\rho \sigma \alpha \beta}$ (the identity
depends only on the Weyl weight of the scalar it is applied to).
As expected, $W_{\mu\nu}^{(i)}(0)$ recover the results in (\ref{W1}) to (\ref{W4})
of $d=4$.

Comparing $W_{\mu \nu}^{(i)}(\epsilon)$ with the four-dimensional case
shows that we have terms similar to those in eqs.\,\eqref{W1}-\eqref{W4}
with the curvatures simply multiplied by the regulator $R^{-\epsilon}$; in addition,
we also  have  terms proportional to $\epsilon$ arising from the variation
of $R^{-\epsilon}$. The latter consists of the operator $- \hat \nabla_{\nu} \hat \nabla_{\mu}
+ R_{\mu \nu} + g_{\mu \nu} \hat \Box $ acting on the original term in the
Lagrangian multiplied by $R^{-1}$ (from  the derivative acting on  $R^{-\epsilon}$
when varying the action).

The  expressions of $W_{\mu\nu}^{(i)}(\epsilon)$ suggest
the emergence (at loop level) of (apparently non-perturbative)
higher dimensional terms suppressed
by powers of $R$. This is similar to quantum  scale invariance in flat
space when such higher dimensional terms emerge at quantum level,
suppressed by powers of dilaton (acting as regulator
and replaced here by $R$) \cite{SI}.

Next, the equation of motion of the dilatation gauge field is
$B_\mu(\epsilon)=0$ with
\medskip
\begin{align}
  B_\mu^{(1)}(\epsilon) & = 4(d-1) \hat \nabla_\mu R^{1- \epsilon}
  - 2\epsilon(d-1) \hat \nabla_\mu \left(R^{-\epsilon-1} R^2 \right)\, ,\\
  B_\mu^{(2)} (\epsilon) & = 2 (d-2) \hat \nabla^\nu (R^{-\epsilon} R_{\mu \nu} )
  + 2 \hat \nabla_\mu R^{1- \epsilon}
  - 2 \epsilon(d-1) \hat \nabla_\mu (R^{-\epsilon-1} R_{\rho \sigma} R^{\rho \sigma})\, ,\\
  B_\mu^{(3)}(\epsilon) & = - 4 \hat \nabla^\nu (R^{-\epsilon} F_{\nu \mu} )
  - 2 \epsilon (d-1)\hat \nabla_\mu (R^{-\epsilon-1} F_{\rho \sigma} F^{\rho \sigma})\, ,\\
  B_\mu^{(4)}(\epsilon) & = 8 \hat \nabla^\nu (R^{-\epsilon} R_{\mu \nu})
  - 2 \epsilon (d-1) \hat \nabla_\mu (R^{-\epsilon-1}
  R_{\rho \sigma \alpha \beta} R^{\rho \sigma \alpha \beta}) \, .
\end{align}

\medskip\noindent
Similar to the case of  $W_{\mu \nu}^{(i)}(\epsilon)$, the same structure
is found for  the correction terms proportional to $\epsilon$: they are
proportional to the operator $\hat \nabla_\mu$ acting on the original
term in the Lagrangian multiplied by $R^{-1}$. 

In arbitrary $d$  dimensions the Euler term, while it remains Weyl gauge covariant,
it is no longer a total derivative and it contributes to the equations of motion.
In the basis of operators  appearing under the integrals of $I_i$, it can be expressed as
\medskip
\begin{equation}
  G = R^2 - 4R_{\mu \nu} R^{\mu \nu}
  + R_{\mu \nu \rho \sigma} R^{\mu \nu \rho \sigma} + 2(d-2)(d-3) F_{\mu \nu} F^{\mu \nu}\, .
\end{equation}

\medskip\noindent
One can check that the equations of motion corresponding to
the linear combination above are non-trivial  in $d\neq 4$. 

\subsection{ Conservation laws}

We can now consider the conservation laws of Weyl gravity in $d$ dimensions.
Calculations are rather long but one can
show that the following identities derived in four dimensions
are also valid in $d$ dimensions:
\medskip
\begin{align}
  \hat \nabla^\mu W_{\mu \nu}^{(i)}(\epsilon) =
  \frac12 F^{\mu}{}_\nu B_\mu^{(i)}(\epsilon)\, ,
  && \hat \nabla^\mu B_\mu^{(i)}(\epsilon) = 2 \tr\, W_{\mu\nu}^{(i)}(\epsilon) \, .
\end{align}

\medskip\noindent
Further, we  have that eq.\,\eqref{divW-2} is  also true in
$d$ dimensions, that is
\medskip
\bea
\hat \nabla^\mu W_{\mu \nu}^{(i)}(\epsilon)
= \mathring \nabla^\mu W_{\mu \nu}^{(i)}(\epsilon) - \w_\nu \, 
g^{\rho \lambda} W_{\rho \lambda}^{(i)}\, .
\eea

\medskip
Therefore  we have the same gauge covariant and Riemannian conservation
law in  $d$ dimensions: if $\w_\mu$ is onshell i.e. $B_\mu(\epsilon)=0$ then 
\medskip
\begin{align}
  \tr\, W_{\mu\nu}(\epsilon) = 0 \qquad \textrm{and}
  \qquad \hat \nabla^\mu W_{\mu \nu}(\epsilon)
  = \mathring \nabla^\mu W_{\mu \nu} (\epsilon)= 0 \, .
\end{align}

\medskip\noindent
For completeness we provide  the expression of 
 $\tr\, W_{\mu\nu}^{(i)}(\epsilon)\propto \hat\nabla^\mu B_\mu^{(i)}(\epsilon)$
\begin{align}
  \tr\, W^{(1)}_{\mu \nu}(\epsilon) & =2(d-1) \hat \Box R^{1-\epsilon}
  - \epsilon(d-1) \hat \Box \left(R^{-\epsilon-1} R^2 \right)\, , \label{trace-d1} \\
  \tr\, W^{(2)}_{\mu \nu}(\epsilon) & = (d-2) \hat \nabla^\mu \hat \nabla^\nu(R^{-\epsilon} R_{\mu \nu})
  + \hat \Box R^{1-\epsilon} -  \epsilon(d-1) \hat \Box (R^{-\epsilon-1} R_{\mu \nu} R^{\mu \nu})\, ,
  \label{trace-d2} \\
  \tr\, W^{(3)}_{\mu \nu}(\epsilon) & = -2\epsilon(d-1) \hat \Box (R^{-\epsilon-1} F_{\mu \nu} F^{\mu \nu})\, ,
  \label{trace-d3}\\
  \tr\, W^{(4)}_{\mu \nu}(\epsilon) & = 4 \hat \nabla^\mu \hat \nabla^\nu (R^{-\epsilon} R_{\mu \nu})
  -  \epsilon (d-1) \hat \Box (R^{-\epsilon-1} R_{\mu \nu \rho \sigma} R^{\mu \nu \rho \sigma})\, .
\label{trace-d4}
\end{align}

\medskip
Consider next the current conservation in $d$ dimensions.
The equation of motion of the dilatation gauge field is a
generalisation of (\ref{def-current}) and has the form
\begin{equation}
4\,\alpha_3 \hat \nabla^\mu (R^{-\epsilon} F_{\mu \nu}) = j_\nu\, ,
\label{current2}
\end{equation}
where
\begin{equation}
  j^\nu = 4(d-1)\alpha_1 \hat \nabla^\nu R^{1- \epsilon}
  - 8(d-3)\alpha_4 G^{\mu \nu} \hat \nabla_\mu R^{-\epsilon}
  - 2\epsilon(d-1) \hat \nabla^\nu (R^{-1} \mathcal L) \, .
\end{equation}

\medskip\noindent
Here $\mathcal L$ denotes the original Lagrangian density of (\ref{action-d2});
$\alpha_3$ is related to  $a_1, \ldots, a_4$ according to \eqref{a-alpha-d}. Note that
the current is modified compared to the $d=4$ case, eq.(\ref{current}).
Above  we  used the identity
\medskip
\begin{equation}
  \hat \nabla^\mu (R^{-\epsilon} R_{\mu \nu}) =
  \frac12 \hat \nabla_\nu R^{1- \epsilon}
  + G_{\mu \nu} \hat \nabla^\mu R^{-\epsilon} \, ,
\end{equation}
where $G_{\mu \nu}$ is the conformal geometry version of the Einstein tensor satisfying
\medskip
\begin{align}
  G_{\mu \nu}& \equiv R_{\mu \nu}- \frac12 g_{\mu \nu} R\, , &
  \hat \nabla^\mu G_{\mu \nu}& = 0\, , & \hat \nabla^\nu G_{\mu \nu} &
  = - (d-2) \hat \nabla^\nu F_{\nu \mu}  \, .
\end{align}

\medskip
As for $d=4$, the trace of $W_{\mu \nu}(\epsilon)$
is related to the divergence of the dilatation current
\medskip
\begin{equation}
 2\,\tr\, W_{\mu\nu}(\epsilon) =\hat \nabla_\alpha j^\alpha  \, .
\end{equation}

\medskip
The conservation of the current (for $\w_\mu$ onshell, $B_\mu(\epsilon)=0$)
follows %
from (\ref{current2})
\begin{align}
  \hat \nabla_\nu j^\nu = \mathring \nabla_\nu j^\nu = 0 \, .
\end{align}

\medskip
In conclusion,  as in the four dimensional case, the dilatation current
is conserved with respect to both  the  Weyl gauge covariant derivative
and Riemannian derivative.  Therefore,  the conservation
laws of Weyl quadratic gravity are valid in arbitrary dimension,
with the dilatation current receiving extra contributions relative to
the case of $d=4$, while respecting  Weyl gauge invariance of the action
in $d$ dimensions. These results are consistent with diffeomorphism
and gauged diffeomorphism invariance which hold in arbitrary dimensions
(see Section \ref{sec-diffeo}).

\section{Conclusions}

In this work we considered the most general (quadratic) action associated to Weyl
conformal geometry  regarded as a gauge theory of the Weyl group.  This
action defines  Weyl quadratic gravity. For this general action
we computed the equations of motion and conservation laws in $d=4$ and then
in arbitrary $d$ dimensions, in a manifestly Weyl gauge covariant way.

We first reviewed in a systematic approach the Weyl gauge covariant
formulation of  Weyl conformal geometry, in which this geometry is automatically
{\it metric} ($\hat\nabla_\mu g_{\alpha\beta}\!=\!0$).
This is a {\it physical} formulation of conformal geometry
regarded as a gauge theory of the Weyl group. There are two other
mathematically equivalent formulations, but they do {\it  not}
guarantee   physical results because they do not have manifest
Weyl gauge covariance.
These are (a): the ``standard'' formulation that has vectorial non-metricity
but is torsion-free, used for over a century,
and (b): a formulation that is metric but has vectorial torsion, that
was discovered recently.  Formulations (a) and (b) are actually related
by a projective transformation  and correspond to different linear
combinations of the Weyl group generators. While (a) and (b)
correspond to an {\it affine} geometry (with an associated Weyl connection),
the physical Weyl gauge covariant formulation does {\it not} correspond to an
affine geometry! This is because it must account for the various
Weyl charges of the fields present and then no connection
$\hat \Gamma$ can be associated to the Weyl gauge covariant derivative
operator $\hat\nabla$. With the geometry being metric
($\hat\nabla_\mu g_{\alpha\beta}=0$),  operator $\hat\nabla$ is for Weyl geometry
  what the derivative $\mathring\nabla$ is for the Riemannian geometry, and can be
  used accordingly. Weyl conformal geometry can then be seen,
  in the gauge covariant formalism, as a  covariantised version
of Riemannian geometry with respect to the dilatations symmetry.

Using this Weyl gauge covariant formalism, we derived the equations of
motion for the most  general quadratic gravity action associated to conformal geometry,
regarded as a (perturbative) gauge theory. Each term in the action has Weyl gauge invariance
and the equations of motion have manifest Weyl gauge covariance.
These results were first obtained for $d=4$ dimensions.
{We were able to generalise these results} 
to arbitrary $d$ dimensions, while respecting this gauge
symmetry. This was possible by considering in $d$ dimensions  the same ($d=4$) action
with an analytical continuation that respects Weyl gauge invariance.
This is realised by a geometric ``regularisation'' enforced by the (covariant)
scalar curvature which replaces  the usual DR subtraction scale $\mu$ (introduced
to keep the couplings dimensionless), so $\mu^{-\epsilon}\ra R^{-\epsilon}$.
This analytical continuation  can also be  derived by an even more general
 Weyl-DBI action in $d$ dimensions,
which in a leading order expansion recovers, up to non-perturbative terms,
 the   action considered here in $d$ dimensions.
This procedure maintained Weyl gauge invariance of the action in $d$
dimensions and the Weyl gauge covariance of the equations of motion.
This is an important result.
The equations of motion in $d$ dimensions suggest the emergence at quantum level
of new (apparently non-perturbative)  higher dimensional Weyl gauge invariant operators
suppressed by powers of the scalar curvature.

We found the conservation laws for general Weyl quadratic gravity in the
conformal geometry covariant formulation, and then also in the equivalent
Riemannian picture.
The energy-momentum tensor of this theory is conserved
with respect to the gauge covariant derivative
$\hat\nabla$ of Weyl conformal geometry but also with respect to
$\mathring\nabla$ operator of the Riemannian picture.
This is an interesting result that is a consequence of having, separately,
standard diffeomorphism invariance (and dilatation invariance)
in a Riemannian picture,  versus gauged covariant diffeomorphisms in conformal geometry.
The same conservation law applies to the Weyl gauge symmetry (dilatation) current.
 This is also a consequence of the dilatations gauge parameter $\lambda_D(x)$
having vanishing Weyl charge. These symmetry arguments are 
independent of the number of dimensions, hence  these conservation laws
apply in arbitrary $d$ dimensions.

The  equations of motions and conservation laws obtained here
in both $d=4$ and in arbitrary $d$ dimensions, in a manifestly
Weyl gauge covariant formalism, can immediately be used for physical
applications. They also  underline the importance of this formalism
for conformal geometry as the gauge theory beyond Einstein-Hilbert gravity
and SM.


\bigskip
\begin{center}
 --------------------------------------------------
  \end{center}

\section*{Appendix}

\def\theequation{A-\arabic{equation}}
\def\thesubsection{A.}
\setcounter{equation}{0}
\def\thefigure{A-\arabic{figure}}
\def\thelabel{A}

\subsection{Conventions}\label{A}

A metric connection $\Gamma$ and a torsionless non-metric connection
$\tilde \Gamma$ can be decomposed in terms of the Levi-Civita $\mathring \Gamma$
and contorsion $K$ or disformation $S$ tensors, respectively
\medskip
\begin{align}
  \Gamma_{\mu \nu}^\rho &= \mathring \Gamma_{\mu \nu}^\rho
  - K_{\mu \nu}{}^\rho\, , & \tilde \Gamma_{\mu \nu}^\rho &
  = \mathring \Gamma_{\mu \nu}^\rho + S_{\mu \nu}{}^\rho \, .
\end{align}

\medskip\noindent
The contorsion $K$ and the disformation $S$ are related to the torsion $T$
and non-metricity $Q$
\begin{align}
  K_{\mu \nu}{}^\rho &= -\frac12 \left(T_{\mu \nu}{}^\rho + T^\rho{}_{\mu \nu}
  - T_\nu{}^\rho{}_\mu \right) \, , & T_{\mu \nu}{}^\rho & = -K_{\mu \nu}{}^\rho + K_{\nu \mu}{}^\rho\, , \\
  S_{\mu \nu}{}^\rho & = -\frac12 \left(Q_{\mu \nu}{}^\rho + Q_\nu{}^\rho{}_\mu
  - Q^\rho{}_{\mu \nu} \right)\, , & Q_{\mu \nu}{}^\rho & = - S_{\mu \nu}{}^\rho - S_{\mu}{}^\rho{}_\nu\, .
\end{align}

\medskip\noindent
The usual definitions of torsion and non-metricity apply
\medskip
\begin{align}
T_{\mu \nu}{}^\rho &= 2 \Gamma_{[\mu \nu]}^\rho\, , & \tilde \nabla_\mu g_{\nu \rho} = Q_{\mu \nu \rho} \, ,
\end{align}

\medskip\noindent
with $\tilde \nabla$ the covariant derivative associated to $\tilde \Gamma$.
Useful integration by parts formulae are
\medskip
\begin{align}\label{a5}
  \int d^d x\, \sqrt{g}\,  \nabla_\mu J^\mu & = \int d^d x \,
  \partial_\mu \left(\sqrt{g} J^\mu \right) - \int d^dx \, \sqrt{g}\, K_{\nu \mu}{}^\nu J^\mu \, ,\\
  \int d^d x\, \sqrt{g}\,  \tilde \nabla_\mu J^\mu & =
  \int d^d x \, \partial_\mu \left(\sqrt{g} J^\mu \right)
  + \int d^dx \, \sqrt{g}\, S_{\nu \mu}{}^\nu J^\mu \, ,
\label{a6}\end{align}

\medskip\noindent
for the two connections. Here $\nabla$ is the operator
associated to $\Gamma$. The traces of the tensors on the right hand side can
also be written as
$K_{\nu \mu}{}^\nu = -T_{\nu \mu}{}^\nu$ and $S_{\nu \mu}{}^\nu = -\tfrac12 Q_{\mu \nu}{}^\nu$
and thus the extra terms depend only on vectorial torsion and semi-metricity.

In the particular case of Weyl quadratic gravity in the equivalent formulations,
metric with vectorial torsion and non-metric but torsion-free, one has
\medskip
\begin{align}
  \Gamma_{\mu \nu}^\rho &= \mathring \Gamma_{\mu \nu}^\rho + \delta_\mu^\rho \w_\nu
  - g_{\mu \nu}\w^\rho \, ,\\
  \tilde \Gamma_{\mu \nu}^\rho & = \mathring \Gamma_{\mu \nu}^\rho
  + \delta_\mu^\rho \w_\nu  + \delta_\nu^\rho \w_\mu - g_{\mu \nu}\w^\rho \, ,
\end{align}

\medskip\noindent
as we saw in the text. Then the corresponding tensors are
\medskip
\begin{align}
  T_{\mu \nu}{}^\rho & = \delta_\mu^\rho \w_\nu - \delta_\nu^\rho \w_\mu\, , & K_{\mu \nu}{}^\rho &
  = g_{\mu \nu} \w^\rho - \delta_\mu^\rho \w_\nu \, , \label{weyl-torsion}\\
  Q_{\mu \nu \rho} & = - 2 \w_\mu g_{\nu \rho}\, , & S_{\mu \nu}{}^\rho & = \delta_\mu^\rho \w_\nu
  + \delta_\nu^\rho \w_\mu - g_{\mu \nu}\w^\rho \, .\label{weyl-non-metricity}
\end{align}
%

\vspace{1cm}
\def\theequation{B-\arabic{equation}}
\def\thesubsection{B.}
\setcounter{equation}{0}
\def\thefigure{B-\arabic{figure}}
\def\thelabel{B}

\subsection{Tangent space formalism}\label{B}

We mentioned throughout the paper that Weyl quadratic gravity is a true gauge
theory of the Weyl group. This can be seen explicitly in the tangent space
formulation of the theory which we review briefly in this appendix. This was discussed at
length in \cite{CDA,CDA2}.

As for any gauge theory, the starting point is the algebra of the group we
want to gauge which, in our case, is given by
\begin{equation}
  \begin{split}
  \label{gauge-algebra}
  &  [P_{a}, M_{bc}] =  \eta_{a b} P_{c} - \eta_{ac}P_a\, , \qquad   [D,P_a] = P_a \, ,
  \qquad [P_a, P_b] = 0, \qquad    [D, M_{ab}] = 0\, , \\[5pt]
& [M_{ab}, M_{cd}] = \eta_{a c}M_{d b} - \eta_{bc}M_{da} - \eta_{ad} M_{cb} + \eta_{bd} M_{ca}\, ,
\end{split}
\end{equation}
where $M^{ab}$ are the generators of the (local) Lorentz group, $P^a$ the generators
of translations and $D$ the generator of dilatations. In a pure gauge theory, where
no matter fields are considered, the only fields in the theory are the gauge fields
associated with the generators of the algebra. In our case we introduce the
spin connection $\Om_\mu{}^{ab}$, vielbein $e_\mu^a$ and Weyl gauge field
of dilatations, $\w_\mu$. The transformations of these fields
as dictated by the algebra are
\begin{align}
  \delta_\epsilon e_\mu^a & = - \partial_\mu \xi^a - \xi_b \Om_\mu{}^{ab}
  - \xi^a \w_\mu + \lambda^a{}_b e_\mu^b + \lambda_D e_\mu^a \, ,\label{var-1p}\\
  \delta_\epsilon \Om_{\mu}{}^{ab} & =- \partial_\mu \lambda^{ab}
  - 2 \lambda^{[a}{}_c \Om_\mu{}^{b]c} \, , \label{var-2p}\\
   \delta_\epsilon \w_\mu & = - \partial_\mu \lambda_D \, . \label{var-3p}
\end{align}
Note that the algebra  above is acting on the tangent
space, hence the explicit tangent space indices on the generators. On  the other
hand we want to interpret our results in space-time. There is no problem to
``uplift'' dilatations and local Lorentz transformations to space-time via the
vielbein and the vielbein postulate. However, translations on the tangent space
need more work. We would like to be able to interpret such translations on the
tangent space as inducing general coordinate transformations in space-time
\begin{equation}
  x^\mu \to x^\mu + \xi^a e_a^\mu \equiv x^\mu + \xi^\mu \, .
\end{equation}
This, however, is in obvious contradiction with the transformations of the fields
on the tangent space shown above. More exactly, the fields $\Om_\mu{}^{ab}$
and $\w_\mu$ are inert under translations even if they  have a
space-time index. To understand how to modify the transformations for these
fields we look at the vielbein which has a non-trivial transformation under
translations on the tangent space. Together with the constraint $R_{\mu \nu}(P^a) = 0$,
discussed later on, the transformation of the vielbein becomes
\cite{CDA,CDA2}
\begin{equation}
  \delta e_\mu^a = - \xi^\rho \partial_\rho e_\mu^a - \partial_\mu \xi^\rho e_\rho^a
  - \xi^\rho \omega_\rho{}^a{}_b e_\mu^b - \xi^\rho \w_\rho e_\mu^a \, . \label{xivar}
\end{equation}
The first two terms are just the ordinary Lie derivative $\mathcal L_\xi$ acting
on the space-time co-vector $e_\mu^a$, the third term is a Lorentz transformation
with parameter $\lambda^a{}_b = - \xi^\rho \Om_\rho{}^a{}_b$, while the last term is a gauge dilatation
\eqref{var-1p} with parameter $\lambda_D = -\xi^\rho \w_\rho$ acting on the vielbein.
We call this transformation a gauge covariant diffeomorphism or covariant general
coordinate transformation.  Therefore
if we want our theory to be invariant under the transformation of the vielbein
above we need to supplement general coordinate transformations of the fields
with the corresponding Lorentz transformation with parameter $- \xi^\rho \Om_\rho{}^a{}_b$
and  dilatation with parameter $-\xi^\rho \w_\rho$. For the spin
connection $\Om_\mu{}^{ab}$, given that it is invariant under dilatations
and taking into account its transformation under Lorentz rotations \eqref{var-2p} one obtains
\begin{equation}
\begin{split}
  \delta \Om_\mu{}^{ab} &= - \xi^\nu \partial_\nu \omega_\mu{}^{ab}
  - \omega_\nu{}^{ab} \partial_\mu \xi^\nu + \partial_\mu(\xi^\nu \omega_\nu{}^{ab})
  - \xi^\nu \omega_\nu{}^a{}_c \omega_\mu{}^{cb}- \xi^\nu \omega_\nu{}^b{}_c \omega_\mu{}^{ac}\\
  & = \xi^\nu R^{ab}{}_{\mu \nu}\, ,
\end{split}
\end{equation}
while for the Weyl field we find
\begin{equation}
  \delta \w_\mu = - \xi^\rho \partial_\rho \w_\mu - \partial_\mu \xi^\rho \w_\rho
  + \partial_\mu( \xi^\rho \w_\rho) = F_{\mu \rho}\, \xi^\rho \, ,
\end{equation}
which are explicitly covariant under dilatations. The same happens with the
transformation of the vielbein \eqref{xivar} which can be recast in the following way
\begin{equation}
  \delta e_\mu^a =- \xi^\rho \hat D_\rho e_\mu^a - \hat D_\mu \xi^\rho e_\rho^a\, .
\end{equation}
Furthermore, this induces a covariant transformation of the metric
\begin{equation}
  \delta g_{\mu \nu} = - \hat \nabla_\mu \xi_\nu - \hat \nabla_\nu \xi_\mu \, .
\end{equation}
Therefore we shall say that the theory is invariant under {\it covariant general
coordinate transformations}, a fact which is crucial in understanding the covariant
conservation laws presented in the main text.

In order to construct actions invariant under the symmetries of the theory we need
to construct covariant objects out of the fields at our disposal. These are in
general the curvatures which according to the gauge algebra are defined as 
\begin{equation}
  \label{fieldstrengths}
  \begin{aligned}
  R_{\mu \nu} (P^a) &= 2 D_{[\mu} e_{\nu]}^a +2 \w_{[\mu} e_{\nu]}^a,
                     \\
R_{\mu\nu}(M^{ab}) & =  \partial_\mu \Om_\nu{}^{ab} -
  \partial_\nu \Om_\mu{}^{ab} + \Om_\mu{}^a{}_c \Om_\nu{}^{cb}
  - \Om_\nu{}^a{}_c \Om_\mu{}^{cb} \equiv R^{ab}{}_{\mu \nu} \, , \\
R_{\mu \nu}(D) & =\partial_\mu \w_\nu - \partial_\nu \w_\mu \equiv  F_{\mu \nu} \, ,
  \end{aligned}
\end{equation}
and are guaranteed to transform covariantly under dilatations
\begin{equation}
  \label{fs-transforms}
    \delta_D R_{\mu \nu}(P^a) =\lambda_D R_{\mu \nu}(P^a)\, , \quad
    \delta_D R_{\mu \nu}(M^{ab}) =0\, , \quad
    \delta_D R_{\mu \nu}(D)  = 0\, .
\end{equation}

In passing we mentioned that (covariant) general coordinate transformations are
obtained after imposing the constraint $R_{\mu \nu}(P^{a}) = 0$. This also makes it
possible to express the spin connection in terms of vielbein and the dilatation
gauge field. Therefore, like in usual Riemannian geometry, the spin connection
is not an independent field, but can be expressed in terms of the other fields
of the theory, in our case the vielbein $e_\mu^a$ and the Weyl field $\w_\mu$
\begin{equation}
  \label{bsol}
  \Om_\mu{}^{ab} = \mathring{\Om}_\mu{}^{ab} +2 e_\mu^{[a} e^{b]\nu} \w_\nu \, ,
\end{equation}
where $\mathring \Om_\mu{}^{ab}$ is the unique torsion-free spin connection
corresponding to the Levi-Civita connection and is given by
\begin{equation}
  \label{spinLC}
  \mathring \Om_\mu{}^{ab} = 2e^{\nu[a} \partial_{[\mu} e_{\nu]}^{b]}
  - e^{\nu[a}e^{b]\sigma}e_{\mu c} \partial_\nu e_\sigma^c \, .
\end{equation}
From a Riemannian perspective the spin connection in \eqref{bsol} has torsion,
but as it was shown in \cite{CDA2} in Weyl quadratic gravity this is not a meaningful
statement as a redefinition of the generators can change the solution for the
connection to a non-metric connection. In particular, we can define the new generators as
\begin{equation}
  \label{Mtilde}
  \tilde M^{ab} = M^{ab} + \tfrac1d D \eta^{ab} \, .
\end{equation}
The associated gauge field, $\tilde \Om_\mu{}^{ab}$ is no longer antisymmetric
in the indices $a, b$ and therefore does not preserve the metric
\begin{equation}
  \tilde D_\mu \eta ^{ab} = 2 \tilde \Om_{\mu}{}^{cd} \eta_{cd} \eta^{ab} \, .
\end{equation}
To fix this issue and show that neither torsion, nor non-metricity are physical
a fully gauge covariant derivative can be constructed as
\begin{equation}
  \label{Dhat}
  \hat D_\mu = D_\mu + q\, \w_\mu = \tilde D_\mu + \tilde q\,  \w_\mu \, ,
\end{equation}
where $q$ and $\tilde q$ denote the tangent space and space-time charges
respectively of the objects on which the derivative acts. The metric is preserved
by this derivative while the torsion which can naively be read from \eqref{bsol}
will be shown below not to appear in the commutator of such derivatives.
Physically, this is the only correct derivative to work with, as it allows us to
construct gauge invariant quantities which are the only ones relevant for physical
observables. The derivatives $D_\mu$ and $\tilde D_\mu$ fail to produce (dilatation)
covariant results when acting on objects which have non-vanishing Weyl charge on
the tangent space and in space-time respectively.

\def\theequation{C-\arabic{equation}}
\def\thesubsection{C.}
\setcounter{equation}{0}
\def\thefigure{C-\arabic{figure}}
\def\thelabel{C}

\subsection{Space-time interpretation}\label{C}

The covariant derivative introduced in \eqref{Dhat} can be ``uplifted''
to space-time using the relation
\begin{equation}
\hat \nabla_\mu V^\nu = e_a^\nu \hat D_\mu V^a \, ,
\end{equation}
which can be deduced from the vielbein postulate
\begin{equation}
  \label{postulate}
  0 = \hat D_\mu e_\nu^a - \tilde \Gamma_{\mu \nu}^\rho e_\rho^a = \partial_\mu e_\nu^a
  + \Om_\mu{}^a{}_b \; e_\nu^b - \Gamma_{\mu \nu}^\rho e_\rho^a \, .
\end{equation}
We can therefore define the full covariant derivative in space-time as
\begin{equation}
  \hat \nabla_\mu X = \nabla_\mu X + q_X \w_\mu X = \tilde \nabla_\mu X + \tilde q_X \w_\mu X\, ,
  \qquad X\equiv X^{\mu_1 \mu_2...}_{\nu_1\nu_2...}\, ,
\end{equation}
where $\nabla_\mu$ and $\tilde \nabla_\mu$ denote usual covariant derivatives with
the affine connections $\Gamma$ and $\tilde \Gamma$ respectively, while $q_X$
and $\tilde q_X$ denote the tangent space and space-time charges respectively.
$X$ denotes a general tensor, but in explicit calculations care must be taken
when raising and lowering indices as the derivative $\tilde \nabla_\mu$ does not
preserve the metric and therefore the charge $\tilde q_X$ has to be modified accordingly.

A further comment is in order. In general, the charges above reflect the way
the tensor $X$ transforms under Weyl gauge transformations, \eqref{gen-transf}.
An exception in this sense is given by the vielbein, which, under transformation
\eqref{gauge-transformation} changes as
\begin{equation}
  e_\mu^a \mapsto e^{\lambda_D} e_\mu^a \, ,
\end{equation}
and its covariant derivatives are given by
\begin{align}
  \hat D_\mu e_\nu^a  &= D_\mu e_\nu^a + \w_\mu e_\nu^a = \tilde D_\mu e_\nu^a\, ,
  \label{viel-1} \\
  \hat \nabla_\mu e_\nu^a & = \nabla_\mu e_\nu^a = \tilde \nabla_\mu e_\nu^a + \w_\mu e_\nu^a\, .
  \label{viel-2}
\end{align}
From above one can see that the Weyl charge $+1$ of the vielbein
is associated either to the tangent space index when $\hat D$ acts or to
the space-time index when $\hat \nabla$ acts on it, and hence calling $q$ or $\tilde q$
tangent space or space-time charges for objects with ``mixed'' indices
depends on which derivative we are considering.
The roles of $q$ and $\tilde q$ are reversed in the case of
the vielbein when passing from one derivative to the other.

The curvatures  can be obtained from the commutator of two $\hat \nabla$ derivatives
\begin{equation}
  \label{hatcommp}
  \big[ \hat \nabla_\mu , \hat\nabla_\nu \big] V^\rho = R^\rho{}_{\sigma \mu \nu} V^\sigma 
  + q_X F_{\mu \nu} V_\rho \, .
\end{equation}
Note that in the above commutator no torsion term is present and this is just the
standard gauge theory formula where all the field strengths have to appear on the
right hand side of the commutator. This supports our claim before that also torsion
(which can be naively read off from \eqref{bsol}) is not physical.

The curvature tensor defined above does not have the same symmetries as the
Riemannian counterpart, but obeys similar Bianchi identities.
The first Bianchi identity reads
\begin{equation}
  \label{IBI}
  R_{\sigma [\mu \nu\rho]} = - g_{\sigma [\mu} F_{\nu \rho]} \, ,
\end{equation}
while the second Bianchi identity, when written in terms of $\hat \nabla$
is similar to  Riemannian case
\begin{equation}
  \label{IIBI}
  \hat \nabla_{[\mu} R^{\tau \sigma}{}_{\nu \rho]} = 0 \, .
\end{equation}

\medskip\noindent
When deriving the equations of motion for Weyl quadratic gravity, one of the most powerful
application of the covariant derivative $\hat \nabla_\mu$ is  that
integration by parts can be performed in the same way as in the Riemannian case.
Simple algebraic manipulations lead to the general formula below,
with $\tilde q_{J^\mu}$ the space-time charge of the current $J^\mu$ (with upper index)
\medskip
\begin{equation}
  \int d^d x\, \sqrt{g}\,  \hat \nabla_\mu J^\mu  =
  \int d^d x \, \partial_\mu \left(\sqrt{g} J^\mu \right)
  + (d+\tilde q_{J^\mu}) \int d^dx \, \sqrt{g}\, \w_\mu J^\mu \, .
\end{equation}

\medskip\noindent
This simplifies further for $q_{J^\mu}=-d$.
  For comparison, see also (\ref{a5}), (\ref{a6}) using $\nabla$ or $\tilde\nabla$.

Next, let us define a vector $J^\mu \equiv X^{\mu \nu_1 \ldots \nu_p} Y_{\nu_1 \ldots \nu_p}$, then
with tensors $X$ and $Y$ both covariant under (\ref{gauge-transformation}) and
$\tilde q_{J^\mu} \equiv  \tilde q_{X^{\mu \nu_1...\nu_p}} + \tilde q_{Y_{\nu_1...\nu_p}}$, we find
\medskip
\bea
  \int d^d x \, \sqrt{g}\, X^{\mu \nu_1 \ldots \nu_p} \hat \nabla_\mu Y_{\nu_1 \ldots \nu_p} &=&
 \int d^d x \,  \partial_\mu \left(\sqrt{g} J^\mu \right) + (d+\tilde q_J)
 \int d^dx \, \sqrt{g}\,\, \w_\mu J^\mu\nonumber\\
&&
 - \int d^dx \, \sqrt{g} \Big(\hat \nabla_\mu X^{\mu \nu_1 \ldots \nu_p} \Big)  Y_{\nu_1 \ldots \nu_p} \, .
\label{parts0}
\eea

\medskip\noindent
In general we are interested in cases where the left hand side
of (\ref{parts0}) is  Weyl gauge invariant i.e. the Weyl charge of the current is
$\tilde q_{J^\mu} = -d$; in that case
the last relation becomes
\bea
  \int d^d x \, \sqrt{g}\, X^{\mu \nu_1 \ldots \nu_p} \hat \nabla_\mu Y_{\nu_1 \ldots \nu_p}
  =\int d^d x \,  \partial_\mu \left(\sqrt{g} J^\mu \right)
  - \int d^dx \, \sqrt{g} \, \left(\hat \nabla_\mu X^{\mu \nu_1 \ldots \nu_p} \right)
  Y_{\nu_1 \ldots \nu_p} \, . \quad
\eea
  For completeness, we add here the values of
  some Weyl charges of fields, connections, etc
that complement those in the table presented in Section~\ref{2}:
\begin{table}[h!]
  \centering
  \begin{tabular}{c|ccccccccccc} 
    $X$ & $\delta \Gamma_{\mu \nu}^\rho$ & $\delta \tilde \Gamma_{\mu \nu}^\rho$
    & $\delta g_{\mu \nu}$ & $\delta \w_\mu$ & $W_{\mu\nu}$ &  $B_\mu$
    &  $ \xi^\mu $ & $\lambda_D$  & $j_\mu$ & $W_{\mu \nu}(\epsilon)$ & $B_\mu(\epsilon)$
     
    \\[2mm]
    \hline
 $\tilde q_X$ &     0 &  0 & 2 &  0 & -2 & -2 & 0 & 0  & -2 & $-(d-2)$ & $-(d-2)$ 
    \\[2mm]
        $ q_X$ &   -1 & -1 & 0 & -1 & -4 & -3 & 1 & 0  & -3 & $-d$     & $-(d-1)$ 
    \\[2mm]
  \end{tabular} \, .
\end{table}

\def\theequation{D-\arabic{equation}}
\def\thesubsection{D.}
\setcounter{equation}{0}
\def\thefigure{D-\arabic{figure}}
\def\thelabel{D}

\subsection{Useful formulae}\label{D}

Using the commutator \eqref{hatcommp} and the Bianchi identities \eqref{IBI} and
\eqref{IIBI} we find several important formulae in $d$ dimensions
used to  simplify the calculations in  the paper. 

First, expanding the commutator in \eqref{hatcommp} we find the standard
formula for the curvature tensor in Weyl geometry (covariant formulation)
\begin{equation}
  R^\rho{}_\sigma{}_{\mu \nu} = \partial_\mu \Gamma^\rho_{\nu \sigma}
  - \partial_\nu \Gamma^\rho_{\mu \sigma} + \Gamma^\rho_{\mu \tau}
  \Gamma^\tau_{\nu \sigma} - \Gamma^\rho_{\nu \tau} \Gamma^\tau_{\mu \sigma} \, .
  \label{curvature-tensor}
\end{equation}

\medskip\noindent
When expressed in terms of the Riemannian geometry curvature tensor, we have
\begin{equation}
\begin{split}
  R_{ \rho \sigma \mu \nu }&  = \mathring{R}_{ \rho \sigma \mu \nu }
    + \left[g_{\mu \sigma} \mathring{\nabla}_\nu \w_\rho
      - g_{\mu \rho} \mathring{\nabla}_\nu  \w_\sigma
      + g_{\nu \rho}  \mathring{\nabla}_\mu
  \w_\sigma - g_{\nu \sigma}\mathring{\nabla}_\mu \w_\rho  \right] \\
  & + \w^2(g_{\mu \sigma} g_{\nu \rho} - g_{\mu \rho} g_{\nu \sigma})
    + \w_\mu(\w_\rho g_{\nu \sigma} - \w_\sigma g_{\nu \rho})
    + \w_\nu(\w_\sigma g_{\mu \rho} - \w_\rho g_{\mu \sigma})\, .
\end{split} \label{R-Rcirc}
\end{equation}

\medskip\noindent
The tensor $R^\alpha{}_{\sigma\mu\nu}=g^{\alpha\rho} \,R_{\rho\sigma\mu\nu}$
is invariant under dilatations. $R_{\rho\sigma\mu\nu}$  is antisymmetric in the
first and second pair of indices, but lacks the usual pair exchange
symmetry of Riemannian geometry curvature tensor
\begin{equation}
  R_{ \rho \sigma \mu \nu} - R_{ \mu \nu \rho \sigma} = -F_{\mu \rho} g_{\sigma \nu}
  + F_{\mu \sigma} g_{\rho \nu} + F_{\nu \rho} g_{\sigma \mu}
  - F_{\nu \sigma} g_{\rho \mu} \, ,  \label{R-1}
\end{equation}

\medskip\noindent
which is a direct consequence of the first Bianchi identity \eqref{IBI}. Further,
contracting two indices we immediately find 
\begin{equation}
  \label{R-Fd}
  R_{\mu \nu} - R_{\nu \mu} = (d-2)\, F_{\mu \nu} \, .
\end{equation}

\medskip\noindent
One can also define a Weyl tensor $C_{\mu \nu \rho \sigma}$ associated to the curvature tensor
\eqref{curvature-tensor}
\medskip
\begin{equation}\label{D6}
  \!\!\!
  C_{\rho \sigma \mu \nu } = R_{ \rho \sigma \mu \nu}
  + \frac1{d-2} \left(R_{ \sigma \mu} g_{\nu \rho} - R_{ \rho \mu} g_{\nu \sigma}
    + R_{ \rho \nu} g_{\mu \sigma} - R_{ \sigma \nu} g_{\mu \rho} \right) 
     +  R\,\, \frac{(g_{\mu \rho} g_{\nu \sigma} - g_{\mu \sigma} g_{\nu \rho})}{(d-1) (d-2)}
\end{equation}

\medskip\noindent
Using \eqref{R-Rcirc}, one checks
that  the Weyl tensor of Weyl geometry in the gauge covariant formulation 
is equal to its Riemannian geometry counterpart
\medskip
\begin{equation}\label{Cw}
C_{\rho \sigma \mu \nu} = \mathring C_{\rho \sigma \mu \nu} \, .
\end{equation}
Its square can be expressed as
\begin{equation}
\begin{split}\label{weyl-squared}
  C_{\mu \nu \rho \sigma} C^{\mu \nu \rho \sigma} &= R_{\rho \sigma \mu \nu} R^{\rho \sigma \mu \nu}
  -\frac{4}{d-2} R_{\mu \nu} R^{\mu \nu} + \frac{2}{(d-1)(d-2)} R^2 \\
  &= R_{\rho \sigma \mu \nu} R^{ \mu \nu \rho \sigma}-\frac{4}{d-2} R_{\mu \nu} R^{\nu \mu}
  + \frac{2}{(d-1)(d-2)} R^2 \, ,
\end{split}
\end{equation}
where, in order to arrive at the second line above, one uses eqs.\,\eqref{R-1}, \eqref{R-Fd}.

The Euler term $G$ that is convenient to use in the action has the following expressions
\medskip
\begin{equation}\label{D9}
\begin{split}
G&= R^2 - 4 R_{\mu \nu} R^{\nu \mu} + R_{\mu \nu \rho \sigma} R^{\rho \sigma \mu \nu} \\[8pt]
&  = R^2 - 4R_{\mu \nu} R^{\mu \nu}
  + R_{\mu \nu \rho \sigma} R^{\mu \nu \rho \sigma} + 2(d-2)(d-3) F_{\mu \nu} F^{\mu \nu} \, .
\end{split}
\end{equation}

\medskip\noindent
Derivative identities known from Riemannian geometry have a very simple form in Weyl
geometry when the derivative $\hat \nabla_\mu$ is used. In particular,
starting from the second Bianchi identity \eqref{IIBI}, by contracting
with $\delta ^\mu_\tau$ one obtains
\begin{equation}
  \hat \nabla_\mu R^\mu{}_{\sigma \nu \rho} = \hat \nabla_\nu R_{\sigma \rho}
  - \hat \nabla_\rho R_{\sigma \nu} \, .
\end{equation}
Contraction of the above with $g^{\sigma \rho}$ gives
\begin{equation}
  \hat \nabla_\mu R^\mu{}_\nu  = \frac12\hat \nabla_\nu R \, ,
\end{equation}
and further applying $\hat \nabla_\nu$ yields
\begin{equation}
  \hat \nabla_\nu \hat \nabla_\mu R^{\mu \nu} = \frac12 \hat \Box R \, .
\end{equation}
These last relations are again similar to those in Riemannian geometry
with the only difference that the order of the indices is important due to \eqref{R-Fd}.

In the covariant formalism the following formulae are  important in 
calculations of  the equations of motion.
We can apply the  general commutator \eqref{hatcommp} to $R_{\rho \sigma}$ and $F_{\rho \sigma}$
to find
\begin{align}
  \big[\hat \nabla_\mu, \hat \nabla_\nu \big] R_{\rho \sigma} & = - R_{\tau \sigma} R^\tau{}_{\rho \mu \nu}
  - R_{\rho \tau} R^\tau{}_{\sigma \mu \nu} -2 F_{\mu \nu} R_{\rho \sigma}\, ,\\
  \big [\hat \nabla_\mu, \hat \nabla_\nu \big] F_{\rho \sigma} & = - F_{\tau \sigma} R^\tau{}_{\rho \mu \nu}
  - F_{\rho \tau} R^\tau{}_{\sigma \mu \nu} -2 F_{\mu \nu} F_{\rho \sigma} \, ,
\end{align}
where we have used that the tangent space charges of $R_{\mu \nu}$ and $F_{\mu \nu}$
are equal to $-2$. After contracting with the metric and using  identity
\eqref{R-Fd} for $d=4$ we have
\begin{align}
\label{id-four}
\big[\hat \nabla_\mu, \hat \nabla_\nu \big] R^{\mu \nu}
&= \big [\hat \nabla_\mu, \hat \nabla_\nu \big] F^{\mu \nu} = 0\, ,
      & \hat \nabla_\nu \hat \nabla_\mu R^{\mu \nu} &=
      \hat \nabla_\mu \hat \nabla_\nu R^{\mu \nu}=\frac12 \hat \Box R\, .
\end{align}
In $d$ dimensions eqs.(\ref{id-four})  become
\begin{align}
\big [\hat \nabla_\mu, \hat \nabla_\nu \big] R^{\mu \nu} & = \frac12 (d-2)(d-4) F_{\mu \nu}F^{\mu \nu}\, ,\\
\big [\hat \nabla_\mu, \hat \nabla_\nu \big ] F^{\mu \nu} & = (d-4) F_{\mu \nu}F^{\mu \nu} \, .
\end{align}

\medskip\noindent
Inserting a suitable factor $R^{-\epsilon}$ (when  we consider $d= 4-2 \epsilon$)
these relations have a closer resemblance to the those for $d=4$:
\medskip
\begin{align}
\label{id-d}
\big [\hat \nabla_\mu, \hat \nabla_\nu \big] (R^{-\epsilon} R^{\mu \nu})
&= \big [\hat \nabla_\mu, \hat \nabla_\nu \big ]
      (R^{-\epsilon}  F^{\mu \nu}) = 0 \, ,\\
\hat \nabla_\nu \hat \nabla_\mu (R^{-\epsilon}  R^{\mu \nu})
&= \hat \nabla_\mu \hat \nabla_\nu (R^{-\epsilon}  R^{\mu \nu})
=\frac12 \hat \Box R^{1- \epsilon} + \hat \nabla^\nu \left(G_{\mu \nu} \hat \nabla^\mu R^{-\epsilon} \right)\, , 
\end{align}

\medskip\noindent
with $G_{\mu \nu} : = R_{\mu \nu} - \frac12 g_{\mu \nu} R$ being a conformal geometry counterpart
to  Einstein tensor.
Notice that $G_{\mu \nu}$ is not symmetric because $R_{\mu \nu}$ is not symmetric.

Using the Bianchi identities for the covariant derivative $\hat \nabla_\mu$ several
simplifying calculations
are possible for terms with derivatives of the Riemann tensor in Weyl geometry.
For example, in $W^{(2)}_{\mu \nu}$ we have $\hat \nabla_\rho \hat \nabla_\nu R^\rho{}_{(\mu \sigma)}{}^\nu$
which can be shown to become in $d=4$ dimensions
\medskip
\begin{equation}
\begin{split}
  \hat \nabla_\rho \hat \nabla_\nu R^\rho{}_{(\mu \sigma)}{}^\nu &= - \hat \Box R_{(\mu \sigma)}
  + \hat \nabla_\rho \hat \nabla_{(\sigma} R_{\mu)}{}^\rho + \hat \nabla_\nu \hat \nabla_{(\sigma} R^\nu{}_{\mu)}
  - \frac12 \hat \nabla_{(\mu} \hat \nabla_{\sigma)} R \\
  & - R_{\lambda (\sigma \mu) \nu} R^{\lambda \nu} - R_{(\lambda \mu)} R_{(\nu \sigma)} g^{\lambda \nu}
  - \frac12 F_\mu{}^\nu R_{\sigma \nu} - \frac12 F_\sigma{}^\nu R_{(\mu \nu)} \, .
\end{split}
\label{Euler-id}
\end{equation}

\medskip\noindent
Two  formulae in $d=4$ used
to derive the energy-momentum tensor $T_{\mu \nu}(\w)$ associated to the
dilatation gauge field are
\medskip
\begin{equation}
\begin{split}
  \hat \nabla_\nu \hat \nabla_\mu R& = \mathring \nabla_\nu \mathring \nabla_\mu R
  - 3 \left(\w_\mu \mathring \nabla_\nu R + \w_\nu \mathring \nabla_\mu R \right)
  - 2 R \mathring \nabla_\nu \w_\mu\\
&+g_{\nu \mu} \left(\w^\rho \mathring \nabla_\rho R - 2 \w^2 R \right) + 8 \w_\mu \w_\nu R \, ,
\end{split} \label{nabla} 
\end{equation}

\medskip\noindent
and, after contracting with the metric
\medskip
\begin{equation}
  \hat \Box R = \mathring \Box R - 2 \w^\rho \mathring \nabla_\rho R
  - 2 R \mathring \nabla_\rho \w^\rho \, .\label{box}
\end{equation}

\medskip
Identical formulae apply
for $R \rightarrow -\phi^2$ which are used to derive $\tilde T_{\mu \nu}(\w,\phi)$.


\end{document}